\documentclass[11pt,a4paper]{article}

\pdfoutput = 1

\usepackage{jcappub}

\usepackage{bm}

\usepackage{rotating}
\usepackage{multicol}
\usepackage{enumerate}
\usepackage{amssymb}

\newcommand{\lambdabar}{{\hbox{$\lambda$\kern-1.ex\raise+0.45ex\hbox{--}}}}
\DeclareMathAlphabet{\mathpzc}{OT1}{pzc}{m}{it}

\newcommand{\ws}{\textcolor{white}{!}}
\newcommand{\be}{\begin{equation}\centering}
\newcommand{\ee}{\end{equation}}
\newcommand{\bea}{\begin{eqnarray}}
\newcommand{\eea}{\end{eqnarray}}

\newcommand{\fett}[1]{\boldsymbol{#1}}

\newcommand{\dd}{{\rm{d}}}

\newcommand{\myPsi}[1]{\fett{\Psi}(\fett{q}_{#1})}

\newcommand{\intp}{\int \frac{\dd^3 p}{(2\pi)^3}}
\newcommand{\pp}{P_L(k_1)\,P_L(|\fett{k}_2-\fett{p}|)\,P_L(p)}
\newcommand{\ppp}{P_L(k_1)\,P_L(k_2)\,P_L(p)}
\newcommand{\ii}{{\rm{i}}}

\makeatletter
\def\fleq{\@fleqntrue\let\mathindent\@mathmargin \@mathmargin=\leftmargini}
\def\cneq{\@fleqnfalse}
\makeatother 

\begin{document}

\begin{flushright}
{\large \tt 
TTK-12-10}
\end{flushright}

\title{Lagrangian perturbations and the matter bispectrum II: the resummed one-loop correction to the matter bispectrum}

\author{Cornelius~Rampf and Yvonne~Y.~Y.~Wong} 

\affiliation{  
Institut f\"ur Theoretische Teilchenphysik und Kosmologie, RWTH Aachen, D--52056 Aachen, Germany}

\emailAdd{rampf@physik.rwth-aachen.de, yvonne.wong@physik.rwth-aachen.de}

\abstract{This is part two in a series of papers in which we investigate an approach based on Lagrangian perturbation theory (LPT) 
to study the non-linear evolution of the large-scale structure distribution in the universe.
Firstly, we compute the matter bispectrum in real space using LPT up one-loop order, for both Gaussian and non-Gaussian initial conditions. 
In the initial position limit, we find that the one-loop bispectrum computed in this manner is identical to its counterpart
obtained from standard Eulerian perturbation theory (SPT). 
Furthermore, the LPT formalism allows for a simple reorganisation of the perturbative series  corresponding to 
the resummation of an infinite series of perturbations in SPT.   Applying this method, we find
a resummed one-loop bispectrum that compares favourably with results from $N$-body simulations.
We generalise the resummation method also to the computation of the redshift-space  bispectrum up to one loop.}

\maketitle

\section{Introduction}

Current measurements of  the cosmic microwave background  (CMB) anisotropies
and large-scale structure (LSS)  distribution strongly support the validity of the so-called 
$\Lambda$CDM model~\cite{Komatsu:2010fb}.
In this model, quantum fluctuations on an inflaton field set the seeds of primordial curvature perturbations in 
the very early universe.  These perturbations manifest themselves as inhomogeneities in the matter/energy fields, and 
are observed as temperature and polarisation anisotropies in the CMB.  
Subsequent evolution via gravitational instability significantly enhances these initial perturbations, ultimately leading to the formation of 
the cosmic structures we see today.

Although the basic $\Lambda$CDM paradigm has so far been remarkably successful at explaining a host of astrophysical
observations,  it is nonetheless crucial to devise more tests to constrain the model's parameters, to find its boundaries of validity, and to 
obtain more insight into the details of its sub-scenarios.
One interesting issue is whether or not the  primordial curvature perturbations conform to a perfect Gaussian 
distribution.  (Almost) perfect Gaussianity is a hallmark feature of a large class of simple inflation models, namely,  the canonical single-field slow-roll  models~\cite{Maldacena:2002vr}.
Nonetheless, many  other observationally consistent inflation models are capable of producing primordial perturbations that deviate significantly from Gaussianity  (i.e., primordial non-Gaussianity; PNG)~\cite{Bartolo:2004if}.  Therefore, any future measurement or constraint of PNG 
would strongly limit the inflationary model space.

In general, the term ``Gaussianity'' for a random variable $\delta(x)$ can 
be defined in terms of the distribution of complex phases $\phi$
from the Fourier decomposition of an ensemble of random realisations.
A uniform distribution in $[0,  2 \pi]$ indicates a completely randomised, and hence Gaussian, distribution;
any deviation from uniformity  is to be identified 
with ``non-Gaussianity''~\cite{Matsubara:2003te}.  
The much-used two-point statistics, the power spectrum---defined as  
$P(k)\! =\!|\tilde\delta(k)|^2$,
where  $k\!\equiv\!|{\fett k}|$ is the Fourier wavenumber---is not well-suited to study non-Gaussianities, 
because it is independent of the phases of the perturbation variable $\tilde\delta(k)$.
The lowest order statistics that is directly sensitive to the phases is 
the three-point function, or  the bispectrum $B(k_1,k_2,k_3)$, defined via $\langle \tilde\delta(\fett{k}_1)
\tilde\delta(\fett{k}_2) \tilde\delta(\fett{k}_3) \rangle_c = (2 \pi)^3 \delta_D^{(3)}(\fett{k}_1 + \fett{k}_2  +\fett{k}_3) B(k_1,k_2,k_3)$.
Gaussian fluctuations necessarily lead to $B(k_1,k_2,k_3)\!=\!0$, since the sum of the phases 
$\Phi \equiv \phi(\fett{k}_1)+\phi(\fett{k}_2)+\phi(\fett{k}_3)$ must also be uniformly distributed in $[0, 2 \pi]$, so that
the ensemble average $\langle e^{i\Phi}  \rangle$
 evaluates to zero.    Contrastingly, a non-uniform, and hence non-Gaussian, phase distribution always yields a non-zero $B(k_1,k_2,k_3)$.

Although a non-zero bispectrum necessarily indicates the presence of non-Gaussianity, it is important to realise that 
non-Gaussianities arise generically as a result of non-linear coupling between different Fourier modes.
This means that while linear evolution of the primordial curvature perturbations preserves their statistical properties, 
as soon as linear theory fails to describe the evolution of inhomogeneities,
a sizable bispectrum can be expected from non-linear evolution alone, 
even in the absence of PNG.
A particularly  relevant case is the clustering statistics of the present-day LSS distribution;
non-linear evolution of the matter density perturbations at low redshifts necessarily generate non-Gaussianities of its own.  Therefore, 
if we wish to use the LSS bispectrum as a probe of PNG, it is important that we know how to filter out this ``late-time'' contribution.

Fully non-linear evolution of the matter density perturbations is in general not amenable to analytical treatments.
However, if we are merely interested in the mildly non-linear regime---where  linear theory still dominates, and non-linear  evolution
contributes a small correction to the linear solution---then semi-analytic techniques based on solving a set of fluid equations  using
a perturbative expansion generally return reasonable results (e.g.,~\cite{Scoccimarro:2000zr,McDonald:2006hf}).
Non-linear corrections to the LSS power spectrum up to two loops \cite{Crocce:2007dt,Taruya:2009ir,Carlson:2009it} have been computed within the framework of 
standard perturbation theory~(SPT)  in, e.g.,~\cite{Suto:1990wf,Jain:1993jh,Bernardeau:2001qr,Crocce:2005xy,Crocce:2005xz},
while references~\cite{Matsubara:2007wj,Okamura:2011nu}  work within the framework of
 Lagrangian perturbation theory~(LPT).  
The bispectrum, on the other hand, has only been treated using SPT up to one loop~\cite{Scoccimarro:1997st,Sefusatti:2009qh}.

In this work, we compute for the first time the tree-level contribution and the 
one-loop correction to the LSS bispectrum using the framework of Lagrangian perturbation theory.
We use the perturbative solutions up to fourth order from our accompanying paper~\cite{BuchertRampf:2012}, and 
show that, in the so-called initial position limit, the one-loop bispectrum obtained using the LPT formalism is identical to its
SPT counterpart.  Furthermore, an advantage of the LPT formalism is that the perturbative series can be easily reorganised,
so that an infinite series of perturbations in SPT is effectively resummed~\cite{Matsubara:2007wj}.  Generalising this resummation method
to the bispectrum calculation, we find a ``resummed'' one-loop bispectrum that, when compared with a na\"{\i}ve expansion, is generally a better approximation to  the ``exact'' bispectrum extracted from 
$N$-body simulations in the mildly non-linear regime. 
Resummation techniques have been explored in many matter power spectrum calculations~\cite{Scoccimarro:2000zr,Crocce:2005xy,Crocce:2005xz,McDonald:2006hf,Matsubara:2007wj,Pietroni:2008jx,Okamura:2011nu}.  To our knowledge, they have as yet not been applied 
to the computation of  the bispectrum.

The paper is organised as follows.  In section~\ref{sec:formalism} we review 
the formalism of LPT, and write down the general expression for the bispectrum.
We evaluate this expression~in section~\ref{sec:realspace} up to one-loop order,
using both a simple Taylor expansion and the aforementioned resummation technique.
The resulting LPT bispectrum and resummed bispectrum are then compared with 
data from $N$-body simulations.  In section~\ref{sec:redshiftspace}, we generalise the 
resummation technique to the computation of the redshift-space bispectrum up to one-loop order.
Our conclusions can be found  in section~\ref{sec:concl}.
In general we try to keep the technical details in the main text to the minimum necessary for the sake of readability;
the reader will be referred to the appendices for the details of the computations at the appropriate points. Here we highlight especially appendix~\ref{app:pertKernels}, where we report all perturbative kernels in both LPT and SPT up to fourth order.

\section{Formalism\label{sec:formalism}}

We briefly review in this section the concepts of Lagrangian perturbation theory (LPT), and construct the bispectrum using the central object of LPT, the displacement
field $\fett{\Psi}$.  Since  the scales that require non-linear corrections are generally well inside the Hubble horizon, a full general relativistic treatment is not necessary, and 
we work within the Newtonian limit of cosmological perturbation theory~\cite{Baldauf:2011bh}.   For readers wishing to skip to the crux of this work, the starting base of our 
bispectrum calculation is equation~(\ref{noResum}), while equation~(\ref{ready}) shows the reorganised perturbative series that is the starting expression for the resummed bispectrum.

\subsection{Lagrangian perturbation theory\label{sec:lpt}}

In the Lagrangian framework,  the observer follows the trajectories of the individual fluid elements~\cite{Zeldovich:1969sb}, 
where each trajectory is encoded in the time-integrated displacement field $\fett{\Psi}$.
The comoving position $\fett{x}$ of a fluid element at conformal time $\tau$
is then given by its initial Lagrangian 
coordinate $\fett{q}$  plus the displacement field $\fett{\Psi}$ evaluated at the same time:
\be \label{trafo}
  \fett{x}(\fett{q},\tau) = \fett{q}+ \fett{\Psi}(\fett{q},\tau) \,.
\ee
The volume of the specific fluid element generally deforms as a result of gravitationally induced displacement.  
The deformation is encoded in  $\dd^3 x\! =\! J (\fett{q},\tau)\, \dd^3 q$, where the 
Jacobian determinant $J\! = \! \det [ \partial \fett{x} / \partial \fett{q} ]$ is generally a function of $\fett{\Psi}$.
Mass conservation (for a non-relativistic fluid) then leads to the constraint:
\be
\label{eq:masscons}
\rho(\fett{x},\tau)\, \dd^3 x = \overline{\rho}(\tau)\, \dd^3 q \,, \qquad 
\ee 
where $\rho(\fett{x},\tau) \equiv \overline{\rho}(t) \left[1+\delta(\fett{x},\tau) \right] $ is the Eulerian density field with
 density contrast $ \delta(\fett{x},\tau)$, and
$\overline\rho(\tau)$ is the mean mass density.
Note that in writing down the constraint~(\ref{eq:masscons}), we have assumed the 
initial density contrast~$\delta(\fett{q},\tau_{\rm ini})$ at $\fett{q}$ to be negligibly small~\cite{BuchertRampf:2012}.
Simple algebraic manipulations of equation~(\ref{eq:masscons}) then allow us to relate the density 
contrast to the displacement field:
\be
 \label{deltaReal}
  \delta(\fett{x},\tau) = \frac{1}{J(\fett{q},\tau)}-1  \,. 
\ee
In the following, where there is no confusion, we will omit writing out the  time dependence of the dynamical variables.

It remains to write down an equation of motion for $\fett{\Psi}$.  In a non-rotating (Eulerian) frame, the equations of motion for self-gravitating and irrotational dust are
\cite{Buchert:1987xy,Buchert:1989xx,Buchert:1995bq}:
\begin{eqnarray}
 \label{contiEul} &&\frac{\partial \delta}{\partial \tau} +  \fett{\nabla_x} \cdot \left\{ \left[ 1+\delta(\fett{x},\tau) \right] \fett{u}(\fett{x},\tau)\right\}= 0 \,, \\
 \label{Euler} &&
\fett{\nabla_x} \cdot  \left[ \frac{\dd \fett{u}}{\dd \tau}  + {\cal H} (\tau)   \fett{u}(\fett{x}, \tau) \right]= - \frac{3}{2} {\cal H}^2 \delta(\fett{x},\tau), \\
 \label{irrotEul} &&\fett{\nabla_x} \times \fett{u} (\fett{x},\tau)= \fett{0}  \,,
\end{eqnarray}
where $\fett{u} = \dd \fett{x}/ \dd \tau$ is the peculiar velocity, $\dd/\dd \tau = \partial/\partial \tau + \fett{u} \cdot \fett{\nabla_x}$  the convective derivative, and
the conformal Hubble parameter evaluates to ${\cal H} = 2/\tau$  in a matter-dominated flat universe.
The last equation~(\ref{irrotEul}) states the irrotationality of the fluid motion in Eulerian space. 
We call equations~(\ref{contiEul}) to~(\ref{irrotEul})  the Euler-Poisson system.

To turn the Euler-Poisson system into equations of motion for $\fett{\Psi}$, 
we first convert the Eulerian derivative $\fett{\nabla_x}$ to its Lagrangian counterpart using the Jacobian of the coordinate transform,
while formally requiring it to be invertible.\footnote{Lagrangian solutions exist even if the Jacobian is not invertible. However, equivalence between the 
Euler-Poisson system and the Lagrange-Newton system can be established only  if $J \neq 0$~\cite{BuchertRampf:2012}.}
Defining $J_{i,j} := \delta_{ij}+ \Psi_{i,j}$, where the subscript ``$,j$'' denotes a partial derivative with respect to the Lagrangian coordinate $q_j$,  the resulting set of equations contains only one dynamical variable,  
namely, the displacement field $\fett{\Psi}$~\cite{Buchert:1993mp}: 
\begin{align}
 \label{LN1} &\varepsilon_{ilm} \varepsilon_{jpq}J_{p,l}(\fett{\Psi}) J_{q,m}(\fett{\Psi})  \frac{\dd^2 J_{j,i}(\fett{\Psi})}{\dd \eta^2} = \frac{12}{\eta^2} \left[ J(\fett{\Psi}) -1 \right]  \,,   \\
 \label{LN2} &J_{i,n}(\fett{\Psi})\, \varepsilon_{njk} J_{l,j}(\fett{\Psi}) \frac{\dd J_{l,k}(\fett{\Psi})}{\dd \eta} = 0 \,, \hspace{0.86cm} {\rm{with}} \,\,\, J > 0 \,,
\end{align}
where $\dd \eta\! =\! \dd \tau/a$ is the superconformal time, and summation over repeated indices is implied.
We call this closed set of equations the Lagrange-Newton system.

Importantly,  the Eulerian irrotationality condition~(\ref{irrotEul})  gives rise to a set of Lagrangian constraints on the displacement field, i.e.,~equation~(\ref{LN2}); in general, $\fett{\Psi}$ contains both 
a longitudinal and a transverse component, that latter arising from the non-linear transformation between the Eulerian and the Lagrangian frame. \emph{Lagrangian} transverse fields  are a necessary constituent of the total displacement field, as they ensure irrotationality of the fluid motion in the Eulerian frame. This means that both (Lagrangian) longitudinal and (Lagrangian) transverse parts affect equation~(\ref{LN1}), and hence the (Eulerian) longitudinal part of the Euler equation~(\ref{contiEul}) as well. Thus, there is in general \emph{no} decoupling between the longitudinal and the transverse components.

Equations~(\ref{LN1}) and~(\ref{LN2}) can be solved using a perturbative series $ \fett{\Psi} = \sum_{n=1}^{\infty}  \fett{\Psi}^{(n)}$.
The linear ($n \!= \! 1$) solution in the initial position limit%
\footnote{The initial position limit corresponds to the case in which the Poisson equation for the linear displacement potential is evaluated at the initial Lagrangian position; see the accompanying paper~\cite{BuchertRampf:2012}.}
  is simply the Zel'dovich approximation~\cite{Zeldovich:1969sb,Buchert:1992ya,BuchertRampf:2012},
\begin{equation}
\label{eq:linearsolution}
\fett{\nabla_q} \cdot \fett{\Psi}^{(1)}(\fett{q},\tau) = - D(\tau)\, \delta^{(1)}(\fett{q},\tau_0) \equiv - D(\tau) \, \delta_0(\fett{q}),
\end{equation}
where $D(\tau)$ is the linear growth function normalised to unity at $z\!=\!0$,
 and the solution is purely longitudinal. Working with the solution~(\ref{eq:linearsolution}),
it is then easy to build iteratively from equations~(\ref{LN1}) and~(\ref{LN2}) higher order solutions.  Expressed in Fourier space and keeping only the fastest growing mode,
the resulting $n$th order displacement field $\tilde{\fett{\Psi}}^{(n)}(\fett{p}) \equiv {\cal F} [\fett{\Psi}^{(n)}(\fett{q})](\fett{p})$ is then
\begin{align} \label{finish4th}
  {\tilde{\fett{\Psi}}}^{(n)} (\fett{p},\tau) &= -\ii D^n(\tau) \int \frac{\dd^3 p_1 \cdots \dd^3 p_n}{(2\pi)^{3n}} 
  \, (2\pi)^3 \delta_D^{(3)}(\fett{p}_{1\cdots n} -\fett{p} ) \, \nonumber \\
 &\ws \hspace{2cm} \times \fett{S}^{(n)} (\fett{p}_1, \ldots, \fett{p}_n) 
  \, \tilde{\delta}_0 (\fett{p}_1) \cdots \tilde{\delta}_0 (\fett{p}_n) \,,
\end{align}
where  $\fett{p}_{1\cdots n}= \fett{p}_1 + \fett{p}_2 + \cdots + \fett{p}_n$, and~\cite{BuchertRampf:2012,Catelan:1994ze,Matsubara:2007wj}
\begin{eqnarray} \label{Ls}
 \fett{S}^{(1)}(\fett{p}_1) &=& \frac{\fett{p}_1}{p_1^2} \,, \\
 \fett{S}^{(2)}(\fett{p}_1, \fett{p}_2) 
  &=&  \frac{3}{7} \frac{\fett{p}_{12}}{p_{12}^2} \frac{\kappa_2^{(s)}}{p_1^2 p_2^2}\,, \\
 \label{ls3rd}\fett{S}^{(3)}(\fett{p}_1, \fett{p}_2, \fett{p}_3) &=&  \frac{\fett{p}_{123}}{p_{123}^2 \,p_1^2 p_2^2 p_3^2} 
      \left[ \frac{1}{3} \kappa_{3a}^{(s)} -\frac{10}{21} \kappa_{3b}^{(s)}  \right]
  +  \frac 1  7 \frac{\fett{\omega}_{3c}^{(s)}}{p_{123}^2 \,p_1^2 p_2^2 p_3^2} \,,  \\ 
 \fett{S}^{(4)}(\fett{p}_1, \fett{p}_2, \fett{p}_3, \fett{p}_4) 
  &=& \frac{\fett{p}_{1234}}{p_{1234}^2\, p_1^2 p_2^2 p_3^2 p_4^2} 
    \left[  \frac{51}{539} \kappa_{4a}^{(s)} - \frac{13}{154} \kappa_{4b}^{(s)} 
             - \frac{14}{33} \kappa_{4c}^{(s)} 
\label{lastls} +\frac{20}{33} \kappa_{4d}^{(s)} +\frac{1}{11} \kappa_{4e}^{(s)}  \right] \, , \nonumber \\
 &&\qquad +\frac{1}{p_{1234}^2\, p_1^2 p_2^2 p_3^2 p_4^2} 
    \left[ \frac 1 6 \fett{\omega}_{4f}^{(s)} - \frac{5}{21} \fett{\omega}_{4g}^{(s)} +\frac{1}{14} \fett{\omega}_{4h}^{(s)}  \right] \,.
\end{eqnarray}
Here, the symmetrised kernels $\kappa_n^{(s)} \equiv \kappa_n^{(s)} (\fett{p}_1, \ldots, \fett{p}_n)$ and  $\fett{\omega}_{n}^{(s)} \equiv \fett{\omega}_n^{(s)} (\fett{p}_1, \ldots, \fett{p}_n)$ 
represent respectively the longitudinal and the transverse component of $\tilde{\fett{\Psi}}^{(n)}$.  Their exact forms can be found in appendix~\ref{app1}.  
We emphasise again that while the displacement field can be split into purely longitudinal and purely transverse components, 
 longitudinal-transverse mixing does occur generically when $\fett{\Psi}$ is used to reconstruct the density contrast via equation~(\ref{deltaReal}).
As we shall see later, for our particular application, there is a mixing at fourth order between $\fett{\omega}_{3c}^{(s)}$ and  $\fett{S}^{(1)}$
(see also in appendix \ref{appX}).  In more general cases, however,  longitudinal-transverse mixing happens already at third order~\cite{Buchert:1993xz}.

\subsection{Bispectrum in the Lagrangian framework\label{sec:bispec}}

Starting with the Fourier transform of the density contrast~\cite{Taylor:1996ne},
\be \label{delta}
  \tilde{\delta}(\fett{k}) \equiv \int \dd^3 x \,e^{\ii \fett{k}\cdot \fett{x}} \delta(\fett{x}) 
  =  \int \dd^3 q \, e^{\ii \fett{k}\cdot\fett{q}}
 \left( e^{\ii \fett{k}\cdot\fett{\Psi}(\fett{q})} -1  \right) \,,
\ee
where we have used equation~(\ref{deltaReal}) for the last equality,
the bispectrum $B(k_1,\,k_2,\,k_3)$  can be defined as
\be \label{defBi}
  \left\langle \tilde{\delta}(\fett{k}_1) \, \tilde{\delta}(\fett{k}_2) 
 \,  \tilde{\delta}(\fett{k}_3) \right\rangle_{\rm c}  \equiv
   (2\pi)^3  \delta_D^{(3)} (\fett{k}_1+\fett{k}_2+\fett{k}_3 ) \,B(k_1,\,k_2,\,k_3)  \,.
\ee
Here, $k_i  \equiv |\fett{k}_i |$,  the subscript ``c'' denotes the connected piece, and 
 the Dirac delta follows from  statistical homogeneity, so that the three wavevectors~$\fett{k}_{1,2,3}$ 
always form a closed triangle.  The dependence of $B(k_1,k_2,k_3)$ on only the magnitude 
of the wavevectors is a consequence of statistical isotropy.
Physically, this means that the bispectrum depends only on the shape and the size of the triangle formed by 
$\fett{k}_{1,2,3}$, not on the triangle's orientation. 
Keeping this in mind, we can rewrite the bispectrum~(\ref{defBi}) using the last equality in equation~(\ref{delta}) as
\begin{align} \label{noResum}
 B(k_1,\,k_2,\,k_3) = &\int \dd^3 \Delta_{21} \int \dd^3 \Delta_{31} 
 \, e^{\ii \fett{k}_2 \cdot \fett{\Delta}_{21}+ \ii \fett{k}_3 \cdot \fett{\Delta}_{31}}  \nonumber \\
     &\times\left( \Big\langle e^{\ii \fett{k}_2\cdot \left(  \myPsi{2}-\myPsi{1} \right) 
+\ii \fett{k}_3\cdot \left( \myPsi{3}- \myPsi{1} \right)} \Big\rangle_c -1 \right) \,,
\end{align}
where $\fett{\Delta}_{ij} \equiv \fett{q}_i - \fett{q}_j$. 
Note that in deriving equation~({\ref{noResum}), we have used  $\langle \exp\{ \ii \fett{k}\cdot\fett{\Psi} \}\rangle=1$, because of $\langle \tilde\delta(\fett{k}) \rangle =0$.

To evaluate the ensemble average in equation~(\ref{noResum}), we use the cumulant expansion theorem~\cite{Ma,Matsubara:2007wj}, 
\be \label{cum}
 \langle e^{\ii X} \rangle = \exp \left\{ \sum_{N=1}^{\infty} \frac{\ii^N}{N!} 
   \langle X^N \rangle_c  \right\} \,, 
\ee
where, in our case, $X=\fett{k}_2 \cdot \left[  \myPsi{2}-\myPsi{1} \right] 
+\fett{k}_3\cdot \left[ \myPsi{3}- \myPsi{1} \right]$, and $ \langle X^N \rangle_c $
denotes the $N$th cumulant.  Two options are available to us at this stage: 
\begin{enumerate}
\item A brute-force Taylor expansion of $\tilde\delta$ in equation~(\ref{delta}) up to the desired order in $\fett{\Psi}$, which yields $\tilde\delta \approx \tilde{\delta}^{(1)} + \tilde\delta^{(2)}+ \ldots$ for equation~(\ref{defBi}).
\item We generalise the resummation scheme of~\cite{Matsubara:2007wj} to the bispectrum calculation, and reorganise 
the perturbative series {\it before} performing a Taylor expansion up to the desired order in 
$\fett{\Psi}$.
\end{enumerate}
 As we shall see in section~\ref{sec:density},  option~1 in fact leads to the same outcome as standard perturbation theory~(SPT), up to the same order in $\tilde\delta$~(this is true at least up to fourth order).   
 
 To implement option~2, we first rewrite equation~(\ref{cum}) as
\begin{eqnarray}
\label{eq:binomial}
    \exp\Bigg\{ \sum_{N=1}^\infty \frac{\ii^N}{N!} 
  \left\langle X^N   \right\rangle_c  \Bigg\} &=& \exp\left\{ \sum_{N=1}^\infty \frac{\ii^N}{N!} D_N \right\}
  \exp\left\{ \sum_{N=1}^\infty \frac{\ii^N}{N!} \, \sum_{j=1}^{N-1} 
 \left(\begin{array}{c} N \\ j \end{array}\right) E_N \right\}  \,,
\end{eqnarray}
where we have used the binomial theorem, and introduced the quantities
\begin{eqnarray}
\label{eq:dn}
  D_N &\equiv& \left\langle a^N \right\rangle_c
 +\left\langle b^N \right\rangle_c+\left\langle c^N \right\rangle_c,   \\
\label{eq:en} 
  E_N &\equiv& \left\langle a^j b^{N-j} \right\rangle_c  
     +\sum_{k=0}^j   \left(\begin{array}{c} j \\ k \end{array}\right) 
  \left\langle a^k b^{j-k} c^{N-j} \right\rangle_c  \,,
\end{eqnarray}
with the shorthand notation 
$a\! \equiv \! \fett{k}_1 \cdot \myPsi{1}$,  $b \! \equiv \! \fett{k}_2 \cdot \myPsi{2}$, and $c \! \equiv \! \fett{k}_3 \cdot \myPsi{3}$.
Using equation~(\ref{eq:binomial}) and the fact that the $D_N$ term is independent  of $\fett{\Delta}_{ij}$, the bispectrum expression~(\ref{noResum}) now becomes
\begin{align} \label{ready}
 B(k_1,\,k_2,\,k_3) &= \exp\left\{ \sum_{N=2}^\infty \frac{\ii^N}{N!} D_N(\fett{0}) \right\} 
  \, \int \dd^3 \Delta_{21} \int \dd^3 \Delta_{31}  
 e^{\ii \fett{k}_2 \cdot \fett{\Delta}_{21}+ \ii \fett{k}_3 \cdot \fett{\Delta}_{31}} \nonumber  \\
 &\ws\hspace{4.5cm}\times \left( \exp\left\{ \sum_{N=1}^\infty \frac{\ii^N}{N!} \, \sum_{j=1}^{N-1} 
 \left(\begin{array}{c} N \\ j \end{array}\right) E_N \right\}  -1 \right) \,,
\end{align}
where, because of statistical homogeneity, we can choose to evaluate $D_N$ at any single point $\fett{q}$ in space,  e.g., at $\fett{q}=\fett{0}$.
For large separations  $\fett{\Delta}_{ij}$, one expects the cumulants in $D_N$ to have a larger 
contribution to  $B(k_1,\,k_2,\,k_3)$ compared with those in $E_N$~\cite{Matsubara:2007wj}. 
Therefore, we expand perturbatively only the cumulants appearing in $E_N$, 
while the exponential prefactor is kept as it is, with the $D_N$ terms in the exponent evaluated according to 
 the Zel'dovich approximation~\cite{Crocce:2005xy,Zeldovich:1969sb}.   This is thus the ``resummation'' scheme 
of~\cite{Matsubara:2007wj}, and equation~(\ref{ready}) is the starting expression for the so-called resummed bispectrum.

\section{Results I: Clustering in real space}\label{sec:realspace}

Working with the expressions~(\ref{noResum}) and~(\ref{ready}), we now evaluate the bispectrum and the resummed bispectrum respectively 
up to one-loop order.  For readers wishing to skip the technical details, the main result
of this section is equation~(\ref{RES}).  A numerical comparison of our results with an ``exact'' bispectrum extracted from $N$-body simulations
can be found in section~\ref{sec:compNbody}, while the equivalence between the LPT and the SPT bispectra is discussed in section~\ref{sec:density}.

\subsection{N-point correlators of the displacement field}\label{sec:multi}

The terms $D_N$ and $E_N$ in equation~(\ref{ready}) contain cumulants that are functions of the displacement fields $\fett{\Psi}(\fett{q})$ (see equations~(\ref{eq:dn}) and~(\ref{eq:en})).
Since the perturbative solutions of $\fett{\Psi}(\fett{q})$ have been conveniently expressed in terms of their Fourier transforms in equations~(\ref{finish4th}) and~(\ref{Ls}), 
we rewrite the cumulants similarly in Fourier space~\cite{Matsubara:2007wj}:
\begin{align}
\left\langle \tilde{\Psi}_{i_1}(\fett{p}_1) \cdots \tilde{\Psi}_{i_N}(\fett{p}_N)  \right\rangle_c 
  &= (2\pi)^3\, \delta_D^{(3)} (\fett{p}_{1\cdots N}) \, C_{i_1 \cdots i_N} (\fett{p}_1, \ldots, \fett{p}_N) \,. \\
\intertext{Here, $C_{i_1 \cdots i_N} (\fett{p}_1, \ldots, \fett{p}_N)$ is an $N$-point correlator, and its perturbative version is defined via}
 \label{perCor} \left\langle \tilde{\Psi}_{i_1}^{(a_1)}(\fett{p}_1) 
  \cdots \tilde{\Psi}_{i_N}^{(a_N)}(\fett{p}_N)  \right\rangle_c 
   &= (2\pi)^3 \,\delta_D^{(3)} (\fett{p}_{1\cdots N}) 
 \,C_{i_1 \cdots i_N}^{(a_1 \cdots a_N)} (\fett{p}_1, \ldots, \fett{p}_N) \,.  
\end{align}
Note that, unlike the $N$-point functions of the density contrast, these $N$-point correlators are in general complex quantities.

Assuming Gaussian initial conditions, the correlators relevant to our problem are:
\begin{eqnarray} \label{curly}
 \begin{array}{rl}
 C_{ij}(\fett{p}) &= \, C_{ij}^{(11)}(\fett{p}) 
  + C_{ij}^{(22)}(\fett{p}) + C_{ij}^{\left\{(31)\right\}}(\fett{p})\,,  \\
 C_{ijk}(\fett{p}_1, \fett{p}_2, \fett{p}_3) &= \, C_{ijk}^{\left\{(211)\right\}} 
  + C_{ijk}^{(222)}+ C_{ijk}^{\left\{(411)\right\}}  
  + {}_{{\rm \MakeUppercase{ \romannumeral 1}} 
  \oplus {\rm \MakeUppercase{ \romannumeral 2}}}C_{ijk}^{\left\{(321)\right\}} \,,  \\
 C_{ijkl}(\fett{p}_1, \fett{p}_2, \fett{p}_3, \fett{p}_4) &= 
  \, C_{ijkl}^{\left\{(1122)\right\}} + C_{ijkl}^{\left\{(3111)\right\}} \,,
 \end{array}
\end{eqnarray}
where we have employed a shorthand notation for the 2-point correlator,  
$C_{ij}(\fett{p}) \equiv C_{ij}(\fett{p},-\fett{p})$, and omitted writing out explicitly the dependences of $C_{ij \cdots N}^{\left\{(ab \cdots)\right\}} 
= C_{ij \cdots N}^{\left\{(ab \cdots)\right\}}(\fett{p}_1,\ldots, \fett{p}_N )$. 
The curly brackets in the superscripts denote summation over all possible permutations, 
e.g., $C_{ijk}^{\left\{(211)\right\}} = C_{ijk}^{(211)}+C_{ijk}^{(121)}+C_{ijk}^{(112)}$, while
the notation
${}_{{\rm \MakeUppercase{ \romannumeral 1}} \oplus {\rm \MakeUppercase{ \romannumeral 2}}}C_{ijk} 
\equiv {}_{{\rm \MakeUppercase{ \romannumeral 1}} }C_{ijk}+{}_{{\rm \MakeUppercase{ \romannumeral 2}} }C_{ijk}$ 
indicates two distinct contributions to correlators of the type $C_{ijk}$.
  In case the initial conditions contain PNG, the correlators 
  \begin{eqnarray}
\label{eq:nongauss}
 \begin{array}{rl}
  C_{ij}(\fett{p}) &= C_{ij}^{\left\{(21)\right\}}(\fett{p})\,,   \\
  C_{ijk}(\fett{p}_1, \fett{p}_2, \fett{p}_3) 
  &=  C_{ijk}^{(111)} + {}_{{\rm \MakeUppercase{ \romannumeral 2}}}C_{ijk}^{\left\{(211)\right\}}
  +{}_{{\rm \MakeUppercase{ \romannumeral 1}
  \oplus {\rm \MakeUppercase{ \romannumeral 2}}}}C_{ijk}^{\left\{(221)\right\}} 
  +  {}_{{\rm \MakeUppercase{ \romannumeral 1}}
  \oplus{\rm \MakeUppercase{ \romannumeral 2}}}C_{ijk}^{\left\{(311)\right\}} \,,  \\
  C_{ijkl}(\fett{p}_1, \fett{p}_2, \fett{p}_3, \fett{p}_4) 
  &= C_{ijkl}^{(1111)} +  C_{ijkl}^{\left\{(2111)\right\}} \,,
 \end{array}
\end{eqnarray}
must be considered in addition.

We can now evaluate equations~(\ref{curly}) and~(\ref{eq:nongauss}) using the perturbative solutions of $\tilde{\Psi}$ given in equations~(\ref{finish4th}) and~(\ref{Ls}),
and express each correlator in terms of the linear power spectrum
$P_L(k,z)\!\!=\!\!D^2(z) P_0(k)$, where $D(z)$ is the linear growth function at redshift $z$, normalised such that $D(z\!=\!0)\!=\!1$, and 
  $P_0(k)$ is the present-day linear power spectrum.
 In the case of non-Gaussian initial conditions, the additional correlators~(\ref{eq:nongauss}) will also depend
 on the present-day linear bispectrum $B_0$ and trispectrum $T_0$.
The exact forms of all correlators in equations~(\ref{curly}) and~(\ref{eq:nongauss})
can be found in appendix \ref{app3}, while in appendix \ref{app2} we give the relations of $B_0$ and $T_0$ to their primordial
counterparts $B_\Phi$ and $T_\Phi$ predicted by inflation.

\subsection{The LPT bispectrum and the resummed bispectrum}\label{sec:results}

We now use the perturbative correlators from section~\ref{sec:multi}  to evaluate the (unresummed) bispectrum~(\ref{noResum}) up to one-loop order.
We begin by separating the perturbative expansion of $B$ 
into a tree-level part $B^{(0)}$ and a one-loop contribution $B^{(1)}$, i.e.,
\be
  B = B^{(0)} + B^{(1)} + \ldots \,,
\ee
which are themselves split respectively into a Gaussian and a non-Gaussian contribution, 
\be \label{Bseries}
  B^{(0)} = B_{211} + B_0 \,, \qquad  B^{(1)} = B_{{\rm Gaussian}}^{(1)} + B_{{\rm PNG}}^{(1)} \,,
\ee
where $B_{211}$ is the tree-level bispectrum with Gaussian initial conditions, and $B_0$ is present 
only in the case of non-zero PNG.
The exact form of $B_0$ is reported in appendix~\ref{app2}.   Here, we explicitly derive $B_{211}$ in order to demonstrate how the formalism works. 

The bracketed term $(\langle \cdots \rangle_c-1)$ in the expression~(\ref{noResum}) can be Taylor-expanded using the cumulant expansion theorem~(\ref{cum}) as
\begin{eqnarray} 
\label{mainexpansion}
  &&\left\langle e^{\ii X} \right\rangle_c -1 =
         -\frac{\ii}{6} \langle X^3 \rangle_c + \frac{1}{24} \langle X^4 \rangle_c 
 + \frac{1}{8} \langle X^2 \rangle_c \langle X^2 \rangle_c  \nonumber \\ 
 &&\ws \hspace{2.6cm} 
  + \frac{\ii}{12} \langle X^2  \rangle_c \langle X^3 \rangle_c 
  - \frac{1}{48}\langle X^2 \rangle_c \langle X^2 \rangle_c   \langle X^2 \rangle_c  + {\cal O}(X^7)\,,
\end{eqnarray}
where  $X = \fett{k}_2 \cdot [\myPsi{2} -\myPsi{1}] +  \fett{k}_3 \cdot [\myPsi{3} -\myPsi{1}]$.
Considering only Gaussian initial conditions, the lowest order contribution is
\begin{eqnarray} \label{treelevel}
  &&B_{211} =  -\ii\, k_{1i} k_{2j} k_{3l} \, C_{ijl}^{\left\{(211)\right\}}(\fett{k}_1,\fett{k}_2, \fett{k}_3)
+k_{1i} k_{2j} C_{ij}^{(11)}(\fett{k}_1) \,k_{2l} k_{3m} C_{lm}^{(11)}(\fett{k}_3)  \nonumber \\
 && \ws \hspace{0.6cm}+k_{1i} k_{2j} C_{ij}^{(11)}(\fett{k}_2) \, k_{1l} k_{3m} C_{lm}^{(11)}(\fett{k}_3) 
  +k_{1i} k_{3j} C_{ij}^{(11)}(\fett{k}_1) \, k_{2l} k_{3m} C_{lm}^{(11)}(\fett{k}_2) \,,
\end{eqnarray}
which, using the definitions of the correlators, can be evaluated to give
\be
\label{eq:211spt}
  B_{211} = 2 P_L(k_1) P_L(k_2) F_2^{(s)} (\fett{k}_1, \fett{k}_2) + {\rm{two\,\, permutations}} \,, 
\ee
with  the kernel 
\be
\label{eq:sptkernel2}
  F_{2}^{(s)}(\fett{k}_1, \fett{k}_2) \equiv \frac 5 7 
 + \frac 1 2  \frac{\fett{k}_1 \cdot \fett{k}_2}{k_1 k_2} \left( \frac{k_1}{k_2} + \frac{k_2}{k_1}  \right) 
  + \frac{2}{7} \frac{\left(\fett{k}_1 \cdot \fett{k}_2\right)^2}{k_1^2 k_2^2} \,.
\ee
But equations~(\ref{eq:211spt}) and~(\ref{eq:sptkernel2}) simply reproduce the results of SPT.   Thus, at  tree level, the LPT and SPT bispectra are identical.

Similar expansions up to higher orders in $\fett{\Psi}$ lead to the one-loop expressions
\renewcommand{\arraystretch}{1.3}
\begin{eqnarray} \label{oneloop}
 \begin{array}{ll}
  B_{{\rm Gaussian}}^{(1)} &=  B_{411 \oplus 123 \oplus 222} + B_{1122 \oplus 1113} 
 + B_{{\rm{xx}} \otimes {\rm{yy}}} +B_{11 \otimes 211}  - B_{11 \otimes 11 \otimes 11} \,,   \\
  B_{{\rm PNG}}^{(1)} &= B_{112 \oplus 122 \oplus 113 }
 + B_{1111 \oplus 1112} + B_{11 \otimes 12} + B_{11 \otimes 111} \,,
 \end{array}
\end{eqnarray}
where explicit forms of the constituent terms are given in appendix~\ref{app4}. 
The $\oplus$ symbol denotes a grouping of  several terms with a common origin in
 the Taylor expansion, e.g., the constituents of
$B_{411 \oplus 123 \oplus 222} \equiv B_{411}
+ {}_{{\rm \MakeUppercase{ \romannumeral 1}} \oplus {\rm \MakeUppercase{ \romannumeral 2}}}B_{123}+B_{222}$, with  ${}_{{\rm \MakeUppercase{ \romannumeral 1}} \oplus {\rm \MakeUppercase{ \romannumeral 2}}}B_{123} \equiv {}_{{\rm \MakeUppercase{ \romannumeral 1}}}B_{123}+ {}_{{\rm \MakeUppercase{ \romannumeral 2}}}B_{123}$,
all originate from the $\langle X^3 \rangle_c$ term in equation~(\ref{mainexpansion}).
The $\otimes$ symbol indicates a contribution from the ``product terms'' in  the Taylor expansion, e.g., $B_{11 \otimes 11 \otimes 11}$ 
arises from $\langle X^2 \rangle_c \langle X^2 \rangle_c   \langle X^2 \rangle_c$, and consists of products of three $C_{ij}^{(11)}$'s. 

In order to compare the LPT one-loop expressions with their SPT counterpart, 
we apply a  diagrammatic technique which allows us to regroup the LPT contributions in terms of the diagrams
they produce.  
In our notation, this means rearranging the LPT contributions into the groupings
\begin{eqnarray} 
 \begin{array}{ll}
  B_{{\rm Gaussian}}^{(1)} &=  \tilde{B}_{411} +\tilde{B}_{123} + \tilde{B}_{222} \,,  \\
  B_{{\rm PNG}}^{(1)} &= \tilde{B}_{112} + \tilde{B}_{122} + \tilde{B}_{113} \,.
 \end{array}
\end{eqnarray}
Since SPT  produces the same classes of diagrams, our LPT results can be compared with standard SPT results  
on a diagram-to-diagram basis.
The various classes of SPT and LPT diagrams up to one loop are shown in figure~\ref{fig:diagrams}.
In appendix~\ref{app5} we demonstrate how to construct diagrams 
in LPT, and report the regrouped terms $\tilde B$.   Suffice to say, all regrouped terms in LPT agree with their SPT counterparts:
\be
   B_{\rm LPT}^{(0)} = B_{\rm SPT}^{(0)}\,, \qquad  B_{\rm LPT}^{(1)} = B_{\rm SPT}^{(1)}\,.
\ee
Since the expressions are identical, henceforth we shall omit the subscripts ``LPT'' and ``SPT''.

\begin{figure}[t]


{\centering

\includegraphics[width=\textwidth]{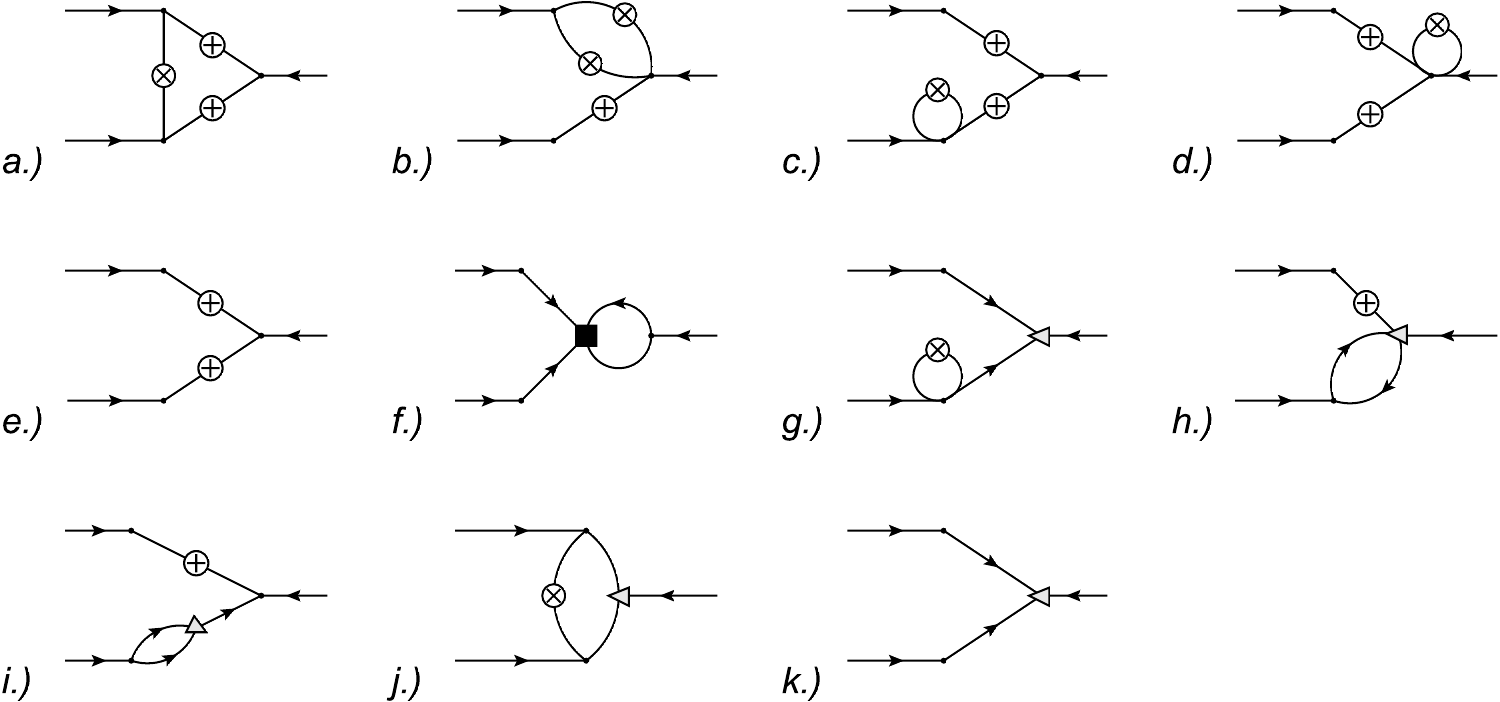}

}

\caption{Diagrams in SPT and LPT. The symbols $\oplus$, $\Delta$ and $\blacksquare$ 
denote respectively a linear power spectrum, linear bispectrum, and linear trispectrum.
For Gaussian initial conditions, the contributing diagrams are (a) $B_{222}$,  (b) ${}_{{\rm \MakeUppercase{ \romannumeral 1}}}B_{321}$, 
(c) ${}_{{\rm \MakeUppercase{ \romannumeral 2}}}B_{321}$, 
(d) $B_{411}$, and  (e) $B_{211}$. 
For non-Gaussian initial conditions, the additional contributions are
(f)
${}_{{\rm \MakeUppercase{ \romannumeral 2}}}B_{211}$, 
(g)
${}_{{\rm \MakeUppercase{ \romannumeral 1}}}B_{113}$,
(h) ${}_{{\rm \MakeUppercase{ \romannumeral 2}}}B_{113}$, 
(i) ${}_{{\rm \MakeUppercase{ \romannumeral 1}}}B_{122}$, 
(j) ${}_{{\rm \MakeUppercase{ \romannumeral 2}}}B_{122}$, and (k)
$B_{111} \equiv B_0$. As discussed in the main text, the sum of all LPT 
contributions to one specific diagram leads to analytic agreement with SPT.
\label{fig:diagrams}}
\end{figure}

To compute the resummed bispectrum, we expand only those terms appearing inside the $\Delta_{21}$ and $\Delta_{31}$
integrals in equation~(\ref{ready}),  while leaving the exponential prefactor untouched.
Expanding up to one-loop order, we find the resummed bispectrum 
\begin{eqnarray}
  &&B(k_1,k_2,k_3) = \exp\left\{ - \frac{1}{2} \int \frac{\dd^3p}{(2\pi)^3}\,
\left[ k_{1i} k_{1j}+k_{2i} k_{2j}+k_{3i} k_{3j}\right]\, C_{ij}^{(11)}(\fett{p})  \right\} \nonumber \\ 
    &&\qquad \hspace{0.7cm}\times \Bigg[  B^{(0)}(k_1,k_2,k_3) + B^{(1)}(k_1,k_2,k_3) 
   + \frac{1}{2}  B^{(0)}(k_1,k_2,k_3)  \Bigg. \nonumber \\   &&\ws \hspace{3.25cm}  \Bigg.\times 
\int \frac{\dd^3p}{(2\pi)^3}\,\left[ k_{1l} k_{1m}
 +k_{2l} k_{2m}+k_{3l} k_{3m}\right]\, C_{lm}^{(11)}(\fett{p})  \Bigg] \,.
\end{eqnarray}
Given  $C^{(11)}_{i j} (\fett{p}) \!=  \!(p_i p_j/p^4) P_L(p)$ from equation~(\ref{eq:linearCij}),
the angular integration of the first and the last terms can be easily performed, thereby leading to
\begin{eqnarray} \label{RES}
 &&B(k_1,k_2,k_3) = 
 \exp\left\{ - \frac{\left(k_1^2 +k_2^2+k_3^2\right)}{12\pi^2} \int \dd p\, P_L(p)  \right\} \nonumber \\ 
    &&\ws \hspace{3cm}\times \left[  B^{(0)} + B^{(1)} 
 + \frac{\left(k_1^2 +k_2^2+k_3^2\right)}{12\pi^2} B^{(0)}
\int \dd p\, P_L(p)     \right] \,,
\end{eqnarray}
with $B^{(0)}$ and $B^{(1)}$ given in equation~(\ref{Bseries}). Equation~(\ref{RES}) is the main result of this paper.

The resummed \emph{power spectrum} was calculated in~\cite{Matsubara:2007wj}.   Comparing 
it with our resummed  bispectrum, one immediately recognises an overall similarity, especially
in the form of an exponential suppression prefactor.  Analogously, as we show in  section~\ref{sec:compNbody}, 
while equation~(\ref{RES}) generally constitutes a better approximation of the bispectrum in the weakly non-linear regime, 
the highly non-linear regime is dominated by the unphysical damping factor.

\subsection{Comparison with N-body results}\label{sec:compNbody}

In this section we compare our resummed bispectrum~(\ref{RES}) with bispectra extracted from 
$N$-body simulations.   We use the $N$-body results of Sefusatti et al.\;2010~\cite{Sefusatti:2010ee}, read off their figures~1, 3 and 5  using the plot digitiser  \texttt{EasyNData}~\cite{Uwer:2007rs}.   The simulations have been performed in a box of side length~$1600\,h^{-1}$Mpc spanned by a $1024^3$ grid,  for a $\Lambda$CDM cosmology 
with   parameters $h\!=\!0.7$, $\Omega_m\!=\!0.279$, $\Omega_b\!=\!0.0462$, $n_s\!=\!0.96$, 
and a fluctuation amplitude $\sigma_8\!=\!0.81$, with initial conditions  generated at redshift $z_i\!=\!99$ using the Zel'dovich approximation.
For the numerical evaluation of equation~(\ref{RES}), we have written a \texttt{C++} code, wherein the integrals 
are evaluated using the deterministic integration routine \texttt{CUHRE} from the \texttt{CUBA library}~\cite{Hahn:2004fe}. 
The code takes as an input the linear power spectrum calculated with \texttt{CAMB}~\cite{Lewis:1999bs}, and linearly interpolates it  for the purpose of the loop integration.


\begin{figure}[t]

\centering 

\includegraphics[width=0.95\textwidth]{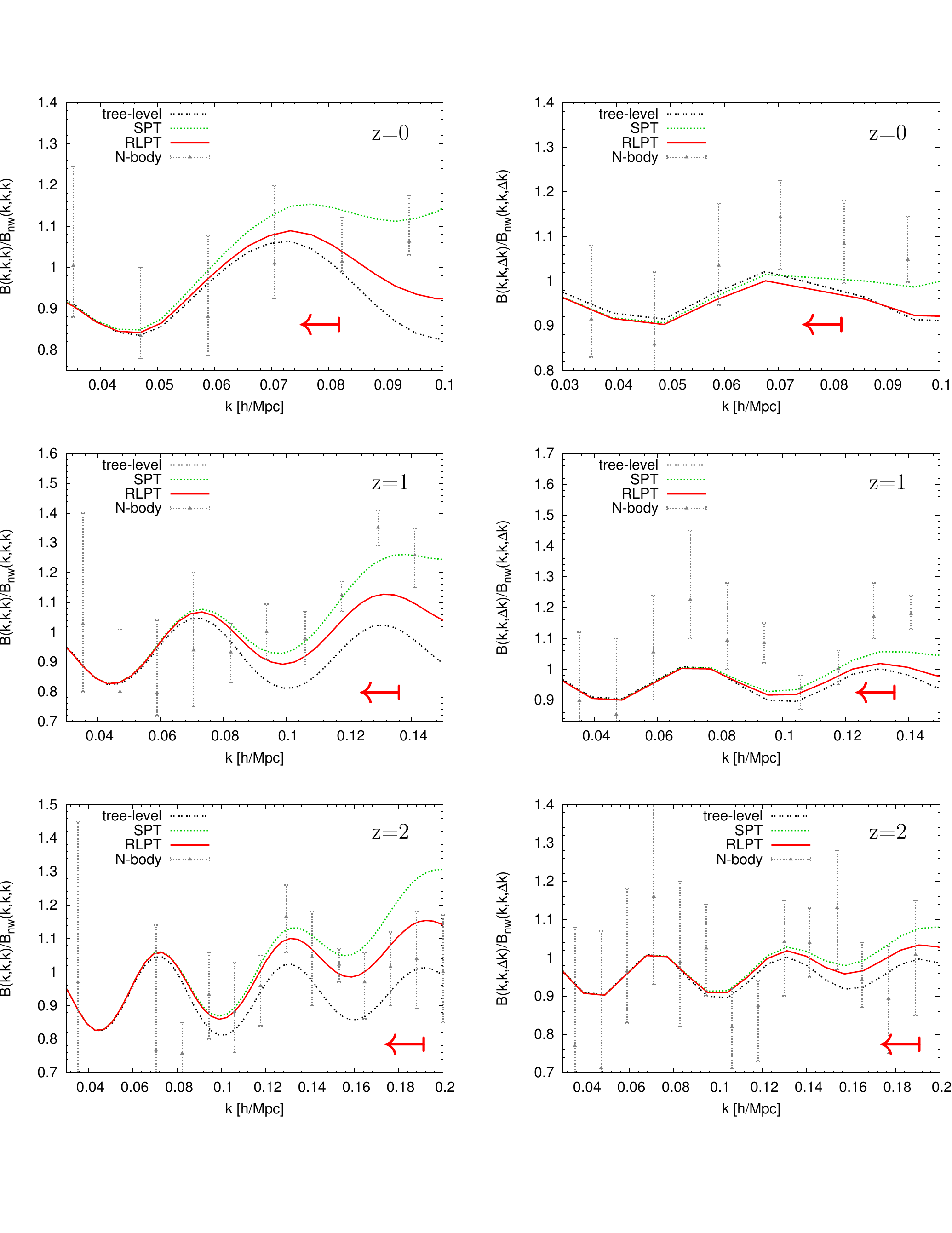}

\vskip-1cm

\caption{Comparison of the one-loop corrected matter bispectra from SPT (green dotted line) and resummed LPT (RLPT; solid red line), with the $N$-body results of \cite{Sefusatti:2010ee}
(data points), at $z\!= \!0$ (\emph{first row}), $z \!= \! 1$ (\emph{second row}), and $z \!= \!2$ (\emph{third row}).   Gaussian initial conditions have been assumed.
{\it Left:}  Bispectrum of the equilateral configuration. 
 {\it Right:} Bispectrum of one squeezed configuration, with $\Delta k \!= \! 0.012 h$/Mpc.   The red arrows indicate the regions of validity for the RLPT results at each redshift.
For reference, we show also the tree-level bispectrum (black dot-dot-space line).  All bispectra have been normalised to the no-wiggle tree-level bispectrum.
\label{fig:resultsZoom}}
\end{figure}

\paragraph{Gaussian initial conditions.} 

Six separate comparisons are shown in figure~\ref{fig:resultsZoom} for the case of Gaussian initial conditions, 
corresponding to bispectra of the equilateral configuration ($k_1\!=\!k_2\!=\!k_3\!\equiv\!k$) 
and a squeezed configuration ($k_1\!=\!k_2\!\equiv\!k$, $k_3\!\equiv\!\Delta k\!=\!0.012\,h/$Mpc), each at redshifts $z\!=\!0,1,2$.
 Each bispectrum displayed has been normalised to the smoothed, 
no-wiggle tree-level bispectrum $B_{\rm nw}$ computed using the transfer functions of~\cite{Eisenstein:1997jh}. In each panel,
the green dotted line denotes the SPT result without resummation, i.e., $B^{(0)}+B^{(1)}$, while the resummed bispectrum (RLPT)
of equation~(\ref{RES}) is represented by the red solid line.

A general deviation of RLPT from the $N$-body results is expected 
in the highly non-linear regime, because the exponential prefactor in~equation (\ref{RES}) has been evaluated only within the Zel'dovich approximation. 
 Therefore, in order to compare our RLPT results with $N$-body data, we must first define a cut-off scale~$k_{\rm eff}$~\cite{Matsubara:2007wj},
\be
  k_{\rm eff} \equiv \alpha \left[ \frac{1}{12 \pi^2} \int \dd p\, P_L(p) \right]^{-\frac 1 2} \,,
\ee
beyond which the exponential damping factor becomes too efficient for our RLPT results to remain physical.  Note that
$k_{\rm eff}$ is time-dependent, and scales with the linear growth factor $D(z)$ as $1/D(z)$ per definition.
The parameter $\alpha$ is a fudge factor that must  be adjusted to the $N$-body data. For $\alpha = 1/3$, we find
good agreement between RLPT and $N$-body data in the $k \leq k_{\rm eff}$ region of validity (indicated 
by the red arrows in figure~\ref{fig:resultsZoom}).  The high redshifts results are especially encouraging:
at $z=2$, RLPT remains compatible with $N$-body data up to $k \sim 0.2\,h$/Mpc, and 
appears to provide a better approximation to the $N$-body bispectrum  than does SPT.

Unfortunately, an error estimation of our approximation with respect to the $N$-body data is in general not possible,
because the simulation errors are often very large.  This is especially so in the case of the squeezed triangle, where the short side
$\Delta k = 0.012\ h$/Mpc is merely a factor of three larger than the fundamental wavenumber of the simulation box ($\sim 0.004\,h$/Mpc).  Sampling 
errors are expected to be large in this instance.


\begin{figure}[t]

\centering 

\includegraphics[width=0.95\textwidth]{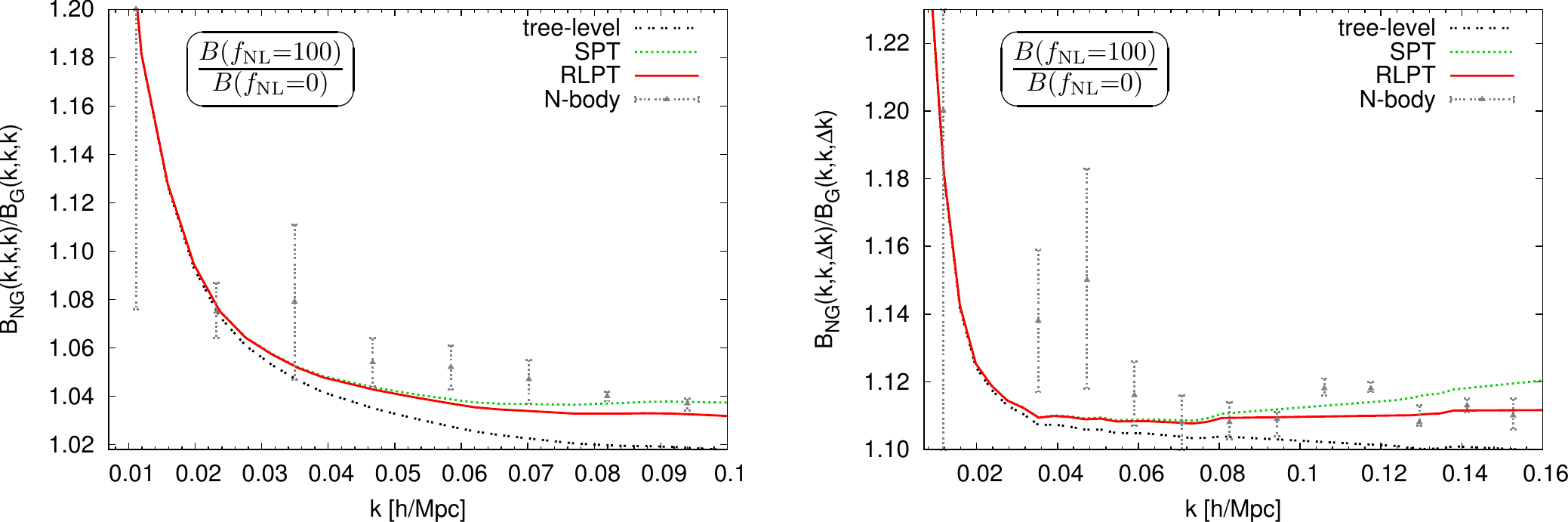}

\caption{Comparison of the one-loop corrected matter bispectra from SPT (green dotted line) and resummed LPT (RLPT; solid red line), with the $N$-body results of \cite{Sefusatti:2010ee}
(data points) at $z\!= \!0$, assuming PNG of the local kind and $f_{\rm NL}\!=\!100$.
 {\it Left:}  Bispectrum of the equilateral configuration.  {\it Right:} Bispectrum of one squeezed configuration, with $\Delta k \!= \! 0.012 h$/Mpc. In each case, the ratio $B(f_{\rm NL}\!= \!100)/ B(f_{\rm NL}\!= \!0)$ is shown.
\label{fig:resultsPNG}}
\end{figure}

\paragraph{Non-gaussian initial conditions.}

Figure~\ref{fig:resultsPNG} shows the case of non-zero PNG of the local type, characterised by $f_{\rm NL}\!=\!100$ (see appendix~\ref{app2} for the definition), 
and its effect on the matter bispectrum at  redshift $z\!=\!0$.  We consider again the equilateral configuration and the same squeezed configuration as above,
computed from one-loop SPT, resummed LPT, and, for reference, tree-level SPT.

Following~\cite{Sefusatti:2010ee}, we plot the ratio $B(f_{\rm NL}\!= \!100)/ B(f_{\rm NL}\!= \!0)$, where $f_{\rm NL}\!=\!0$ corresponds to Gaussian initial conditions.  Results from one-loop SPT are represented by the green dotted line, 
 while the red solid line shows our RLPT calculations.  The reference tree-level results---$B_{211}$ for Gaussian initial conditions, and $B_{211}+B_0$ for the non-Gaussian case---are  
denoted by the black  dot-dot-space line. 
 For convenience we use the non-Gaussian contributions at tree-level and one-loop from 
 figure~1 of~\cite{Sefusatti:2010ee}.

Since the PNG considered here is of the local type, we expect its contribution to the present-day matter bispectrum to be peaked in the squeezed configuration.
Indeed, the contribution due to PNG in the mildly non-linear regime ($k\!\sim\!0.04$--$0.1\,h$/Mpc) is at the 11\% level in the squeezed case, while for the equilateral 
configuration the contribution in the same $k$ range is about 3\%.  The SPT result shows a slight increase in the importance of PNG  at 
 higher $k$-values, while the RLPT result suggests a constant behaviour.  Comparing with $N$-body data,
we find that the $N$-body results tend to overshoot both the RLPT and SPT results.  It is not clear to us at this stage whether this is a problem of the semi-analytic methods, or of the $N$-body simulations.  We note however that one of the main hurdles facing simulations with PNG is the accurate implementation of non-Gaussian initial conditions, and much research in this direction is ongoing (e.g.,~\cite{Liguori:2010hx, Wagner:2011wx, Regan:2011zq}). We conjecture that the generation of initial conditions may
contribute at least partly to the discrepant results.

\subsection{Exact relationship between SPT and LPT}\label{sec:density}

We have demonstrated in section~\ref{sec:results} and the associated appendices that the one-loop bispectra computed from SPT and LPT are identical. Similarly, the equivalence between the SPT and the LPT one-loop power spectra
was previously shown in~\cite{Matsubara:2007wj}. Here, we summarise a result from our accompanying paper~\cite{BuchertRampf:2012}, which shows that LPT and SPT in general return the same results at least up to fourth order in the density contrast $\delta$.

Our starting point is equation~(\ref{delta}).
To prove the equivalence of LPT and SPT,  we expand the LHS of equation~(\ref{delta}) as
\be \label{deltaExp}
 \tilde{\delta} = \tilde{\delta}^{(1)}+\tilde{\delta}^{(2)}+\tilde{\delta}^{(3)}+\tilde{\delta}^{(4)} + \ldots \,,
\ee
and similarly the exponential on the RHS up to the same order in the displacement field $\fett{\Psi}$.
Using explicitly the fastest growing solutions for the $n$th order displacement fields given in equation~(\ref{finish4th}) in the initial position limit, and summing 
up all contributions with $n$ powers of~$\tilde{\delta}_0$, we arrive at
\begin{eqnarray} \label{deltaLag}
  \tilde{\delta}^{(n)}(\fett{k},t) &=& D^n \int \frac{\dd^3 p_1 \cdots \dd^3 p_n}{(2\pi)^{3n}}   \, (2\pi)^3 \delta_D^{(3)}(\fett{p}_{1\cdots n} - \fett{k}) \nonumber \\
  &&\ws \hspace{3cm} \times   X_{n}^{(s)}(\fett{k};\fett{p}_1, \ldots, \fett{p}_n) \, \tilde{\delta}_0 (\fett{p}_1) \cdots \tilde{\delta}_0 (\fett{p}_n) \,,
\end{eqnarray}
where $X_{n}^{(s)}$ is a symmetrised kernel at $n$th order.  Explicit forms can be found in appendix~\ref{appX} for $n\leq 4$. 
The Dirac delta in equation~(\ref{deltaLag}) fixes $\fett{k} = \fett{p}_{1\cdots n}$. A careful examination of the kernels~$X_{n}^{(s)}$
then reveals 
\be
  X_n^{(s)}(\fett{k};\fett{p}_1,\ldots ,\fett{p}_n)\Big|_{\fett{k} = \fett{p}_{1\cdots n}} \Big. = F_n^{(s)}(\fett{p}_1,\cdots ,\fett{p}_n) \,,
\ee
where $F_n^{(s)}$ is simply the $n$th order SPT kernel for the $n$th order density contrast~\cite{Bernardeau:2001qr}, 
shown in a symmetrised form in appendix~\ref{app6} for $n \leq 4$.
The equivalence between $X_{n}^{(s)}$ and $F_n^{(s)}$ has been checked algebraically with a \texttt{Mathematica} code
up to $n=4$~\cite{BuchertRampf:2012}, which is available upon request.\footnote{E-mail to \url{rampf@physik.rwth-aachen.de}.}


\section{Results II: Clustering in redshift space\label{sec:redshiftspace}}

The resummation technique used in this work can be easily generalised to the calculation of the matter bispectrum in redshift space~\cite{Smith:2007sb}. We briefly outline
the computational procedure in the following.  The final expression for the resummed redshift-space bispectrum up to one loop, assuming Gaussian initial conditions, 
can be found in equation~(\ref{redRes}).

\subsection{Density contrast and distorted displacement field in redshift space}\label{sec:red1}

In the so-called plane parallel limit, the comoving distance $\fett{s}$ in redshift space 
is defined as~\cite{Matsubara:2007wj,Kaiser:1987qv}
\be \label{reddis}
  \fett{s} = \fett{x} + \frac{\hat{\fett{z}}  \cdot \fett{v}}{\cal H}\hat{\fett{z}} \equiv \fett{x} 
 - u_z \hat{\fett{z}}  \,,
\ee
where $\fett{x}$ is the comoving distance in real space defined in equation~(\ref{trafo}), 
$\hat{\fett{z}}$ is the line of sight (assuming the observer is fixed on the comoving grid), 
and $\fett{v} \equiv \dd \fett{x}/\dd \tau$ is the 
peculiar velocity of the fluid element. 
Rewriting equation~(\ref{reddis}) as
\be \label{red_trafo}
  \fett{s}(\fett{q},\tau) = \fett{q} + \fett{\Psi}^s(\fett{q},\tau) \,,
\ee
it is easy to see that the distorted displacement field reads
\be \label{red}
  \fett{\Psi}^s =  \fett{\Psi}
 + \frac{\hat{\fett{z}} \cdot  \dd \fett{\Psi}/\dd \tau}{\cal H}  \hat{\fett{z}}\,.
\ee
Mass conservation in redshift and Eulerian space implies~\cite{Scoccimarro:1999ed}
\be
  \overline{\rho} (1+\delta^s(\fett{s}))\, \dd^3 s = \overline{\rho} (1+\delta(\fett{x}))\, \dd^3 x  = \overline{\rho} \, \dd^3 q \,, \qquad
\dd^3 s = J^s \dd^3 q\, ,
\ee
which leads immediately to $\delta^s(\fett{s}) = 1/J^s - 1$, and its Fourier transform
\be
  \label{red_delta}
 \tilde{\delta}^s(\fett{k})  =  \int \dd^3 q \, e^{\ii \fett{k}\cdot\fett{q}} 
  \left( e^{\ii \fett{k}\cdot\fett{\Psi}^s(\fett{q})} -1  \right) \,.
\ee
The redshift-space power spectrum and bispectrum can then be defined as
\begin{align}
   \left\langle \tilde{\delta}^s(\fett{k}) \, \tilde{\delta}^s(\fett{k}') \right\rangle_c 
 &= (2\pi)^3\,  \delta_D^{(3)} (\fett{k}+\fett{k}') \,P^s(\fett{k})  \,, \\
 \left\langle \tilde{\delta}^s(\fett{k}_1) \, \tilde{\delta}^s(\fett{k}_2) 
  \, \tilde{\delta}^s(\fett{k}_3) \right\rangle_c &= 
    (2\pi)^3\,  \delta_D^{(3)} (\fett{k}_1+\fett{k}_2+\fett{k}_3 ) 
  \,B^s(\fett{k}_1,\,\fett{k}_2,\,\fett{k}_3) \,,
\end{align}
on which we can now apply a perturbative procedure analogous to that outlined in section~\ref{sec:bispec} to obtain tractable results.

\subsection{The resummed bispectrum in redshift space}\label{sec:b211Red}

To compute the perturbative correlators in redshift space, we first note that the perturbative kernels $\fett{S}^{(n)}$ given in equation~(\ref{Ls})
are time-independent; the whole time evolution of the displacement field $\fett{\Psi}^{(n)}$ is embedded in a separable linear growth
function $D$ as  $\fett{\Psi}^{(n)}\propto D^n$. Therefore, the time derivative of the $n$th order displacement
field can be written as
\be
 \frac{\dd \fett{\Psi}^{(n)}}{\dd \tau} = n {\cal H} f \fett{\Psi}^{(n)} \,,
\ee
where $f \equiv \dd \ln D / \dd \ln a$ is the logarithmic growth rate. From equation~(\ref{red}) it follows that
\begin{align}
  \fett{\Psi}^{s(n)} &=  \fett{\Psi}^{(n)}+ n f (\hat{\fett{z}} \cdot  \fett{\Psi}^{(n)}) \hat{\fett{z}} \,, \\
\intertext{or, in index notation,}
  \Psi_{i}^{s(n)} &= (\delta_{ij} + n f \hat{z}_i \hat{z}_j) \Psi_j^{(n)} \equiv R_{ij}^{(n)} \Psi_j^{(n)} \,,
\end{align}
where, again, summation over repeated indices is implied.

The tensor $R_{ij}^{(n)}$ deforms the initially isotropic displacement field $\fett{\Psi}^{(n)}$, 
thereby rendering its redshift-space counterpart $\fett{\Psi}^{s(n)}$
 in general direction-dependent.   It follows that 
  the perturbative correlators constructed from the displacement fields must also be similarly distorted, i.e.,
\be
\label{eq:conversion}
  C_{i_1 \cdots i_N}^{s(a_1 \cdots a_N)} (\fett{p}_1, \ldots, \fett{p}_N) 
  = R_{i_1 j_1 }^{(a_1)}  \cdots R_{i_N j_N }^{(a_N)}  C_{j_1 
  \cdots j_N}^{(a_i \cdots a_N)} (\fett{p}_1, \ldots, \fett{p}_N) \,,
\ee
and computing the redshift-space bispectrum becomes simply a matter of replacing all occurrences of   $C_{i_1 \cdots i_N}^{(a_1 \cdots a_N)}$
with   $C_{i_1 \cdots i_N}^{s(a_1 \cdots a_N)}$. 
For example, the tree-level redshift-space bispectrum assuming Gaussian initial conditions is given by
\begin{align}
 &B_{211}^s(\fett{k}_1,\fett{k}_2, \fett{k}_3) =\\
 & \ws \hspace{0.5cm} -\ii \,k_{1i} k_{2j} k_{3l} \, C_{ijl}^{s\left\{(211)\right\}}(\fett{k}_1,\fett{k}_2, \fett{k}_3)
+k_{1i} k_{2j} C_{ij}^{s(11)}(\fett{k}_1) \,k_{2l} k_{3m} C_{lm}^{s(11)}(\fett{k}_3) \nonumber  \\
 & \ws \hspace{0.5cm} +k_{1i} k_{2j} C_{ij}^{s(11)}(\fett{k}_2) \, k_{1l} k_{3m} C_{lm}^{s(11)}(\fett{k}_3) 
 +k_{1i} k_{3j} C_{ij}^{s(11)}(\fett{k}_1) \, k_{2l} k_{3m} C_{lm}^{s(11)}(\fett{k}_2) \,, \nonumber  \\ 
\intertext{cf~the real-space expression~(\ref{treelevel}).  Using the relation~(\ref{eq:conversion}), we obtain,
after some reshuffling, the result}
 &B_{211}^s(\fett{k}_1,\fett{k}_2, \fett{k}_3) =  2\,C(\fett{k}_1,\fett{k}_2, \fett{k}_3) R(\fett{k}_2) R(\fett{k}_3) P_L(k_2) P_L(k_3) \\
  &\ws \hspace{4cm} + \rm{two\,\;cyclic\,\;perms.}\,, \nonumber  \\
\intertext{where} 
 &C(\fett{k}_1,\fett{k}_2, \fett{k}_3) \equiv F_2^{(s)}(\fett{k}_2,\fett{k}_3) 
+ f  \frac{(\fett{k}_1 \cdot \hat{\fett{z}})^2}{k_1^2} G_2^{(s)}(\fett{k}_2, \fett{k}_3)  \\
 & \ws \hspace{0.5cm} - \frac 1 2 f(\fett{k}_1 \cdot \hat{\fett{z}}) 
\left[\frac{(\fett{k}_2\cdot \hat{\fett{z}} )}{k_2^2} 
\left(1+ f \frac{(\fett{k}_3\cdot \hat{\fett{z}})^2}{k_3^2}\right)
+\frac{(\fett{k}_3\cdot \hat{\fett{z}} )}{k_3^2} 
\left(1+ f \frac{(\fett{k}_2\cdot \hat{\fett{z}})^2}{k_2^2}\right) \right]\, ,
  \nonumber \\ 
\label{kaiser}
 &R(\fett{k}_i) \equiv 1+f \frac{(\fett{k}_i \cdot \hat{\fett{z}})^2}{k_i^2}  \,,
\end{align}
with the symmetrised second-order SPT kernels $F_2^{(s)}$ and $G_2^{(s)}$  (see appendix \ref{app6}). 
The redshift-space kernels $R(\fett{k}_i)$ and $C(\fett{k}_1,\fett{k}_2, \fett{k}_3)$ agree exactly with the results from SPT;
in the notation of \cite{Scoccimarro:1999ed}, they are called $Z_1$ and $Z_2$, respectively, with $b\!=\!1$ and $b_2\!=\!0$ 
(i.e., neglecting local galaxy biasing). 

The one-loop correction to the redshift-space bispectrum can be computed in a similar manner.  
 Generalising the procedure to the resummation case, we find the following expression for the resummed 
redshift-space bispectrum up to one loop:
\begin{eqnarray} \label{redRes} 
B^{s}(\fett{k}_1,\fett{k}_2,\fett{k}_3) &=& 
    \exp\left\{ -\frac{\sum_{i=1}^3\left[ k_i^2 +f(f+2) \left( \fett{k}_i \cdot \hat{\fett{z}} \right)^2  \right]  }{12\pi^2} \int \dd p\,P_L(p) \right\}  \\
  &&\quad\times \left[  B^{s(0)} +B^{s(1)}   
 +\frac{\sum_{i=1}^3\left[ k_i^2 
 +f(f+2) \left( \fett{k}_i \cdot \hat{\fett{z}} \right)^2  \right]}{12\pi^2} 
  B^{s(0)} \int \dd p\,P_L(p)  \right] \,. \nonumber 
\end{eqnarray}
Note that, unlike the case of the real-space bispectrum, redshift-space distortions induce for $B^s$ a dependence on the  direction of the constituent wavevectors.
We leave the numerical evaluation of equation~(\ref{redRes}) for a future project.

\section{Conclusions}\label{sec:concl}

In this paper, we have computed the matter bispectrum up to one loop using Lagrangian perturbation theory (LPT).  We find an exact agreement between the bispectrum obtained 
this way and its counterpart from standard Eulerian perturbation theory (SPT), 
as long as the LPT calculation is restricted to the initial position limit~\cite{BuchertRampf:2012}).   
More generally, in this limit, we find that both LPT and SPT predict the same density contrast up to fourth order.  The analytic agreement between these theories up to third order
can be shown by hand, while the equivalence of the fourth-order solutions can only be established algebraically using a \texttt{Mathematica} code (which can be obtained upon request). 
Note that the use of the initial position limit is, to our knowledge, implicit in all LPT calculations in the current literature performed in Fourier space (e.g., \cite{Crocce:2005xy,Matsubara:2007wj,Okamura:2011nu}).  Relaxing this assumption might yield better approximations of the matter $N$-point spectra in the weakly non-linear regime, and should be a promising avenue to study in the future.

We have also generalised a resummation technique, first developed in~\cite{Matsubara:2007wj} to compute a resummed matter power spectrum,
to the matter bispectrum calculation both in real space and in redshift space.
In the mildly non-linear regime where we expect the method to be valid, 
we find good agreement between the ``resummed''  real-space one-loop bispectrum and the ``exact'' bispectrum extracted from the $N$-body simulations of~\cite{Sefusatti:2010ee}
assuming Gaussian initial conditions. Our results for bispectra of the equilateral configuration at high redshifts ($z\!=\!2$) are especially encouraging: within the region of validity, the 
resummed bispectrum (RLPT) generally performs better than its one-loop SPT counterpart in terms of their comparison with the $N$-body data.
The performance of the resummation scheme is more difficult to judge for bispectra of a squeezed configuration.  Here, for a squeezed triangle with a short side
 $\Delta k = 0.012h$/Mpc, we find that the $N$-body bispectra of~\cite{Sefusatti:2010ee} at all considered redshifts have more power in the $k$ region of interest, compared with 
 all semi-analytic approximations studied in this work (i.e., tree-level, one-loop SPT, and resummed one-loop).  However, we note also that the $\Delta k$ value used in the comparison 
 is rather close to the fundamental wavenumber of the simulation box ($\sim 0.004\ h$/Mpc).  This suggests that the $N$-body bispectra in the squeezed limit 
could very well be afflicted by large sampling errors. In the case of non-Gaussian initial conditions of the local variety (with $f_{\rm NL}\!=\!100$), 
 we observe a slight overshoot of the $N$-body results of~\cite{Sefusatti:2010ee} compared to RLPT (and often to SPT as well).   Further studies will be required to understand this discrepancy, but we note for the moment that the crucial issue of generating non-Gaussian initial conditions in $N$-body simulations is highly non-trivial, and research in this direction is ongoing (e.g.,~\cite{Liguori:2010hx, Wagner:2011wx, Regan:2011zq}).

The consideration of  biasing effects in the galaxy clustering statistics (e.g.,~\cite{Jeong:2009vd,Matsubara:2011ck,Matsubara:2008wx}) is beyond the scope of this work.  However, a simple replacement of $f \rightarrow  \beta = f/b$~\cite{Matsubara:2007wj} in our resummed redshift-space bispectrum~(\ref{redRes}) immediately generalises the expression to describe also
 a simple linear biasing scheme with bias parameter $b$. 
  Extending this scenario to non-local biasing schemes is less straightforward, and will be addressed in a future work.


\section*{Acknowledgments} {CR is very grateful to Thomas Buchert for many useful suggestions and an enlightening correspondence.  We thank Takahiko Matsubara and  Emiliano Sefusatti for clarifying their arguments.}


\appendix

\fleq

\section{Perturbative kernels}\label{app:pertKernels}

\subsection{LPT\label{app1}}

The unsymmetrised kernels $\kappa_{n}$ and $\fett{\omega}_n$ in equations~(\ref{Ls}) to~(\ref{lastls}) are given by
 \begin{align}
\label{kernels}
 \kappa_{2} (\fett{p}_1, \fett{p}_2) &= \textcolor{white}{+}
 \frac{1}{2} \left[ p_1^2p_2^2 
  - (\fett{p}_1 \cdot \fett{p}_2)^2  \right] \,, \\ 
 \kappa_{3a} (\fett{p}_1, \fett{p}_2, \fett{p}_3) &= 
   -\frac{1}{6} \left[ p_1^2 p_2^2 p_3^3 - (\fett{p}_1\cdot\fett{p}_2)^2 p_3^2 
 - (\fett{p}_1\cdot\fett{p}_3)^2 p_2^2  - (\fett{p}_2\cdot\fett{p}_3)^2 p_1^2 \right. \nonumber   \\ &\ws \left. \hspace{3cm} +2 (\fett{p}_1\cdot\fett{p}_2) (\fett{p}_1\cdot\fett{p}_3) (\fett{p}_2\cdot\fett{p}_3)   \right] \,, 
\end{align}
\be
 \kappa_{3b} (\fett{p}_1, \fett{p}_2, \fett{p}_3) = 
 -\kappa_{2} (\fett{p}_2, \fett{p}_3) \frac{ \kappa_{2} (\fett{p}_1, \fett{p}_{23})}{p_{23}^2} \,, 
\ee
\be 
\kappa_{4a} (\fett{p}_1, \fett{p}_2, \fett{p}_3, \fett{p}_4) = \textcolor{white}{+}
 \kappa_{2} (\fett{p}_1, \fett{p}_3) \,\kappa_{2} (\fett{p}_2, \fett{p}_4)  \,
 \frac{\kappa_{2} (\fett{p}_{13}, \fett{p}_{24})}{p_{13}^2p_{24}^2 }  \,, 
\ee
\be
 \kappa_{4b} (\fett{p}_1, \fett{p}_2, \fett{p}_3, \fett{p}_4) =  -6\, \kappa_{3a}(\fett{p}_1, \fett{p}_2,\fett{p}_{34}) \, 
 \frac{\kappa_{2} (\fett{p}_3, \fett{p}_4)}{p_{34}^2} \,, 
\ee
\be
 \kappa_{4c} (\fett{p}_1, \fett{p}_2, \fett{p}_3, \fett{p}_4) =  
  -\kappa_{3a} (\fett{p}_2, \fett{p}_3, \fett{p}_{4})  
 \frac{\kappa_{2} (\fett{p}_1, \fett{p}_{234})}{p_{234}^2} \,, 
\ee
\be
 \kappa_{4d} (\fett{p}_1, \fett{p}_2, \fett{p}_3, \fett{p}_4) =  
 - \kappa_{3b} (\fett{p}_2, \fett{p}_3, \fett{p}_4)  
 \frac{\kappa_{2} (\fett{p}_1, \fett{p}_{234})}{p_{234}^2} \,,
\ee
\begin{align}
 \kappa_{4e} (\fett{p}_1, \fett{p}_2, \fett{p}_3, \fett{p}_4) &=  
 -\left[ (\fett{p}_1 \cdot \fett{p}_{2})(\fett{p}_{234} \cdot \fett{p}_{34}) 
 - (\fett{p}_1 \cdot \fett{p}_{34}) (\fett{p}_{234} \cdot \fett{p}_2) \right] \nonumber \\ 
    & \ws \hspace{3.5cm} \times \left( \fett{p}_1 \cdot \fett{p}_{234}  \right) 
  \frac{\fett{p}_2 \cdot \fett{p}_{34}}{p_{234}^2 p_{34}^2} \, \kappa_{2} (\fett{p}_3, \fett{p}_4)  
 \,, 
\end{align}
and
\be
 \fett{\omega}_{3c}(\fett{p}_1,\fett{p}_2,\fett{p}_3) = 
   \left[ \fett{p}_1 \left( \fett{p}_{123} \cdot \fett{p}_{23} \right) - \fett{p}_{23} \left( \fett{p}_{123} \cdot \fett{p}_1 \right)  \right]  \,\left(\fett{p}_1 \cdot \fett{p}_{23} \right)\, \frac{\kappa_2(\fett{p}_2,\fett{p}_3)}{p_{23}^2} \,,
 \ee
\begin{align}   
\fett{\omega}_{4f}(\fett{p}_1,\fett{p}_2,\fett{p}_3,\fett{p}_4)  &= 
   \left[ \fett{p}_1 \left( \fett{p}_{1234} \cdot \fett{p}_{234} \right) - \fett{p}_{234} \left( \fett{p}_{1234} \cdot \fett{p}_1  \right)   \right]\,\left( \fett{p}_1 \cdot \fett{p}_{234} \right) \, \frac{ \kappa_{3a}(\fett{p}_2,\fett{p}_3,\fett{p}_4)}{p_{234}^2}
  \,, \nonumber \\
 \fett{\omega}_{4g}(\fett{p}_1,\fett{p}_2,\fett{p}_3,\fett{p}_4)  &= 
   \left[ \fett{p}_1 \left( \fett{p}_{1234} \cdot \fett{p}_{234} \right) - \fett{p}_{234} \left( \fett{p}_{1234} \cdot \fett{p}_1  \right)   \right] \, \left( \fett{p}_1 \cdot \fett{p}_{234} \right) \, \frac{\kappa_{3b}(\fett{p}_2,\fett{p}_3,\fett{p}_4)}{p_{234}^2} \, ,  \nonumber \\
  \fett{\omega}_{4h}(\fett{p}_1,\fett{p}_2,\fett{p}_3,\fett{p}_4)  &= 
   \left[ \fett{p}_1 \left( \fett{p}_{1234} \cdot \fett{p}_{234} \right) 
     - \fett{p}_{234} \left( \fett{p}_{1234} \cdot \fett{p}_1  \right)   \right] \frac{\fett{p}_1\cdot \omega_{3c}^{(s)}(\fett{p}_2,\fett{p}_3,\fett{p}_4)}{p_{234}^2}\,.
\end{align}
The symmetrisation procedure is then straightforward~\cite{Goroff:1986ep}:
\begin{eqnarray}
  \kappa_n^{(s)}(\fett{p}_1, \ldots, \fett{p}_n) &=& \frac{1}{n!}\sum_{i \in S_n} \kappa_n(\fett{p}_{i(1)}, \ldots, \fett{p}_{i(n)}) \,,  \nonumber \\
  \fett{\omega}_n^{(s)}(\fett{p}_1, \ldots, \fett{p}_n) &=& \frac{1}{n!} \sum_{i \in S_n} \fett{\omega}_n(\fett{p}_{i(1)}, \ldots, \fett{p}_{i(n)}) \,.
\end{eqnarray}
Note that the kernels $\fett{\omega}$ satisfy the condition \cneq
\be
  \label{transverseCondition} \fett{p}_{12\cdots n} \cdot \fett{\omega}_n (\fett{p}_1,\ldots,\fett{p}_n) = 0 \, ,
\ee
indicating transverseness.


\subsection{SPT\label{app6}}

The functions $F_n$ and $G_n$ arising from a perturbative solution of the Eulerian fluid equations~(\ref{contiEul})  to~(\ref{irrotEul})
in an Einstein-de Sitter universe can be obtained using a set of recursion relations~\cite{Bernardeau:2001qr}. 
Explicit expressions up to $n=4$, in their unsymmetrised forms, can be found in, e.g.,~\cite{Goroff:1986ep}.  Here, we give 
the symmetrised versions of these expressions.

\fleq
\begin{itemize}
\item  For $n=2$:
\begin{align}
F_2^{(s)} (\fett{k}_1, \fett{k}_2) &= \frac 5 7 + \frac 1 2 \frac{\fett{k}_1 \cdot \fett{k}_2}{k_1 k_2} \left( \frac{k_1}{k_2}+ \frac{k_2}{k_1} \right) 
  + \frac 2 7 \frac{(\fett{k}_1 \cdot \fett{k}_2)^2}{k_1^2 k_2^2}  \,, \\
G_2^{(s)} (\fett{k}_1, \fett{k}_2) &= \frac 3 7 + \frac 1 2 \frac{\fett{k}_1 \cdot \fett{k}_2}{k_1 k_2} \left( \frac{k_1}{k_2}+ \frac{k_2}{k_1} \right) 
  + \frac 4 7 \frac{(\fett{k}_1 \cdot \fett{k}_2)^2}{k_1^2 k_2^2}  \,.
\end{align}

\item For $n=3$:
\begin{align}
\label{eq:f3}
F_3^{(s)} (\fett{k}_1, \fett{k}_2, \fett{k}_3) &= \frac 1 3 \Bigg\{ \frac{7}{18} \frac{\fett{k}_{123}\cdot\fett{k}_3}{k_3^2}  F_2^{(s)} (\fett{k}_1, \fett{k}_2) \Bigg. \nonumber \\
&\qquad+\left.
\left[ \frac{7}{18} \frac{\fett{k}_{123}\cdot\fett{k}_{12}}{k_{12}^2} + \frac 1 9 k_{123}^2\frac{\fett{k}_{12}\cdot\fett{k}_{3}}{k_{12}^2 k_{3}^2} \right]G_2^{(s)} (\fett{k}_1, \fett{k}_2) \right. \nonumber \\ &\ws \hspace{5cm}+ \Bigg. \rm{2\,cyclic\,permutations} \Bigg\}  \,, 
\end{align}
\begin{align}
G_3^{(s)} (\fett{k}_1, \fett{k}_2, \fett{k}_3) &= \Bigg\{ \frac{1}{18} \frac{\fett{k}_{123}\cdot\fett{k}_3}{k_3^2}  F_2^{(s)} (\fett{k}_1, \fett{k}_2) \Bigg. \nonumber \\
&\qquad+\left.\left[ \frac{1}{18} \frac{\fett{k}_{123}\cdot\fett{k}_{12}}{k_{12}^2} + \frac 1 9 k_{123}^2\frac{\fett{k}_{12}\cdot\fett{k}_{3}}{k_{12}^2 k_{3}^2} \right]G_2^{(s)} (\fett{k}_1, \fett{k}_2) \right. \nonumber \\
 &\ws \hspace{5cm}+ \Bigg. \rm{2\,cyclic\,permutations} \Bigg\}  \,.
\end{align}

\item For $n=4$:
\begin{align}
 &396 \cdot F_4^{(s)}(\fett{k}_1,\fett{k}_2,\fett{k}_3,\fett{k}_4) =
 \left\{ 27 \frac{\fett{k}_{1234}\cdot \fett{k}_1}{k_1^2}  F_3^{(s)} (\fett{k}_2,\fett{k}_3,\fett{k}_4)  \right. \nonumber \\
   &\qquad+ \left. \left[ 27\frac{\fett{k}_{1234}\cdot \fett{k}_{234}}{k_{234}^2} + 6k_{1234}^2 \frac{\fett{k}_1 \cdot \fett{k}_{234}}{k_1^2 k_{234}^2}  \right]G_3^{(s)} (\fett{k}_2,\fett{k}_3,\fett{k}_4) + \rm{3\,cyclic\,permutations} \right\} \nonumber \\
 &\qquad+\left\{ 18 \frac{\fett{k}_{1234}\cdot\fett{k}_{12}}{k_{12}^2} G_2^{(s)}(\fett{k}_1, \fett{k}_2) F_2^{(s)}(\fett{k}_3, \fett{k}_4) + \rm{5\,permutations} \right\} \nonumber \\
 &\qquad+\left\{ 4 k_{1234}^2  \frac{\fett{k}_{12}\cdot \fett{k}_{34}}{k_{12}^2 k_{34}^2} G_2^{(s)}(\fett{k}_1, \fett{k}_2) G_2^{(s)}(\fett{k}_3, \fett{k}_4) + \rm{2\,permutations} \right\}  \,,
\end{align}
\begin{align}
 &396 \cdot G_4^{(s)}(\fett{k}_1,\fett{k}_2,\fett{k}_3,\fett{k}_4) =
 \left\{ 9 \frac{\fett{k}_{1234}\cdot \fett{k}_1}{k_1^2}  F_3^{(s)} (\fett{k}_2,\fett{k}_3,\fett{k}_4)  \right. \nonumber \\
   &\qquad+ \left. \left[ 9\frac{\fett{k}_{1234}\cdot \fett{k}_{234}}{k_{234}^2} + 24 k_{1234}^2 \frac{\fett{k}_1 \cdot \fett{k}_{234}}{k_1^2 k_{234}^2}  \right]G_3^{(s)} (\fett{k}_2,\fett{k}_3,\fett{k}_4) + \rm{3\,cyclic\,permutations} \right\} \nonumber \\
 &\qquad+\left\{ 6 \frac{\fett{k}_{1234}\cdot\fett{k}_{12}}{k_{12}^2} G_2^{(s)}(\fett{k}_1, \fett{k}_2) F_2^{(s)}(\fett{k}_3, \fett{k}_4) + \rm{5\,permutations} \right\} \nonumber \\
 &\qquad+\left\{ 16 k_{1234}^2  \frac{\fett{k}_{12}\cdot \fett{k}_{34}}{k_{12}^2 k_{34}^2} G_2^{(s)}(\fett{k}_1, \fett{k}_2) G_2^{(s)}(\fett{k}_3, \fett{k}_4) + \rm{2\,permutations} \right\}  \,.
\end{align}

\end{itemize}
The advantage of using the above expressions is that the symmetrised $n$th-order kernels have the
property that, in common applications,  partial sums of several wavevectors equate to zero.
For instance,  to compute the loop correction $P_{13}$ to the power spectrum, one integrates over a kernel comprising the function 
$F_{3}^{(s)}(\fett{q}, -\fett{q}, \fett{k})$. Plugging the arguments into equation~(\ref{eq:f3}), 
we see immediately that the second term is proportional to $G_2^{(s)}(\fett{q}, -\fett{q})/(\fett{q}-\fett{q})^2$, where both the numerator and the denominator
evaluate to zero.  A careful treatment shows that this term in fact vanishes as $\varepsilon^2 \cdot 1/\varepsilon$ as $\varepsilon \rightarrow 0$, which is not very straightforward to see from inspecting the corresponding expression in~\cite{Goroff:1986ep}.

\section{Perturbative N-point correlators\label{app3}}

Here we give explicit expressions for the perturbative correlators~(\ref{curly}) and~(\ref{eq:nongauss}),
 computed following the definition~(\ref{perCor}) of $C^{(a_1\ldots a_N)}_{i_1\ldots i_N}$
and using the expressions~(\ref{finish4th}) and~(\ref{Ls}) for $\tilde{\fett{\Psi}}^{(n)}$.

\subsection{Gaussian initial conditions\label{app3.1}}

For Gaussian initial conditions, the correlators~(\ref{curly}) are 
\begin{align}                                                                                                  
\label{eq:linearCij} & C_{ij}^{(11)} (\fett{p}) = \, S_i^{(1)}(\fett{p}) S_j^{(1)}(\fett{p}) P_L(\fett{p}) \,, \\              
& C_{ij}^{(22)} (\fett{p}) = \, 2 \int \frac{\dd^3 p'}{(2\pi)^3}
 S_i^{(2)}(\fett{p}', \fett{p}-\fett{p}') \, S_j^{(2)}(\fett{p}', \fett{p}-\fett{p}') 
 \,P_L(|p-p'|)\,P_L(p') \,, \\ 
&C_{ij}^{(13)} (\fett{p}) = \, 3\, S_i^{(1)}(\fett{p})\,P_L(p) \, \int \frac{\dd^3 p'}{(2\pi)^3}\, S_j^{(3)}(\fett{p}, \fett{p}',- \fett{p}') \, P_L(p') = C_{ji}^{(31)}(\fett{p}) \,,  \\
\label{c211}
& C_{ijk}^{(211)} (\fett{p}_1,\fett{p}_2,\fett{p}_3)  =  \, C_{jki}^{(112)} (\fett{p}_2,\fett{p}_3,\fett{p}_1) 
 = C_{kij}^{(121)} (\fett{p}_3,\fett{p}_1,\fett{p}_2)  \nonumber \\
& \hspace{10mm}=   -2\ii \, S_i^{(2)}(\fett{p}_2, \fett{p}_3) \, S_j^{(1)}(\fett{p}_2) 
 \, P_L(p_2) \,S_k^{(1)}(\fett{p}_3)\,  P_L (p_3) \, , \\
 &C_{ijk}^{(222)} (\fett{p}_1,\fett{p}_2,\fett{p}_3) =
  \,8\ii \int \frac{\dd^3 p'}{(2\pi)^3} S_i^{(2)}(\fett{p}_1+\fett{p}',-\fett{p}')     
  S_j^{(2)}(-\fett{p}_1 -\fett{p}', \fett{p}'-\fett{p}_3) \,\nonumber \\ 
  & \hspace{40mm} \times P_L(|\fett{p}_3-\fett{p}'|) \,P_L(p') 
  S_k^{(2)}(\fett{p}', \fett{p}_3 -\fett{p}') \, P_L(|\fett{p}_1+\fett{p}'|) \,, \\
& C_{ijk}^{(411)} (\fett{p}_1,\fett{p}_2,\fett{p}_3) =C_{jki}^{(114)} (\fett{p}_2,\fett{p}_3,\fett{p}_1) =
 C_{kij}^{(141)} (\fett{p}_3,\fett{p}_1,\fett{p}_2) \nonumber \\
  & \hspace{5mm}= \,12\ii\, S_j^{(1)}(\fett{p}_2)\, 
 P_L (p_2) \, S_k^{(1)}(\fett{p}_3)\, P_L (p_3)   \int \frac{\dd^3 p'}{(2\pi)^3}  
  S_i^{(4)}(-\fett{p}_2,-\fett{p}_3, \fett{p}',-\fett{p}') \,P_L(p')\, , \\
& {}_{\rm \MakeUppercase{ \romannumeral 1}}C_{ijk}^{(123)} (\fett{p}_1,\fett{p}_2,\fett{p}_3) = 
 6\ii \,S_i^{(1)}(\fett{p}_1) \, P_L(p_1) 
     \int \frac{\dd^3 p'}{(2\pi)^3} \, S_j^{(2)}(\fett{p}', \fett{p}_2- \fett{p}') \, \nonumber \\ 
   & \hspace{40mm} \times P_L(|\fett{p}_2-\fett{p}'|) 
 \,S_k^{(3)}(-\fett{p}_1, \fett{p}'-\fett{p}_2 , -\fett{p}') \,P_L(p') \,, \\
&{}_{\rm \MakeUppercase{ \romannumeral 2}}C_{ijk}^{(123)} (\fett{p}_1,\fett{p}_2,\fett{p}_3) =
\,-6\ii \,S_i^{(1)}(\fett{p}_1) \, P_L(p_1) \, S_j^{(2)}(\fett{p}_1, \fett{p}_3) \, P_L(p_3) \nonumber \\
 & \hspace{40mm}  \times \int \frac{\dd^3 p'}{(2\pi)^3} S_k^{(3)}(\fett{p}_3, \fett{p}', -\fett{p}') 
\,P_L(p')  \,,  \\
 &{}_{{\rm \MakeUppercase{ \romannumeral 1}} 
 \oplus {\rm \MakeUppercase{ \romannumeral 2}} }C_{ijk}^{(123)} (\fett{p}_1,\fett{p}_2,\fett{p}_3)  
 \equiv {}_{\rm \MakeUppercase{ \romannumeral 1}} C_{ijk}^{(123)} (\fett{p}_1,\fett{p}_2,\fett{p}_3)  
+ {}_{\rm \MakeUppercase{ \romannumeral 2}} C_{ijk}^{(123)} (\fett{p}_1,\fett{p}_2,\fett{p}_3)  \nonumber \\
 & \hspace{10mm} = {}_{{\rm \MakeUppercase{ \romannumeral 1}} \oplus {\rm \MakeUppercase{ \romannumeral 2}} }C_{jki}^{(231)}
  (\fett{p}_2,\fett{p}_3,\fett{p}_1) = {}_{{\rm \MakeUppercase{ \romannumeral 1}} 
 \oplus {\rm \MakeUppercase{ \romannumeral 2}} }C_{kij}^{(312)} (\fett{p}_3,\fett{p}_1,\fett{p}_2) = {}_{{\rm \MakeUppercase{ \romannumeral 1}} 
  \oplus {\rm \MakeUppercase{ \romannumeral 2}} }C_{ikj}^{(132)} (\fett{p}_1,\fett{p}_3,\fett{p}_2)    \nonumber \\ 
& \hspace{10mm} =\, {}_{{\rm \MakeUppercase{ \romannumeral 1}} 
 \oplus {\rm \MakeUppercase{ \romannumeral 2}} }C_{kji}^{(321)} (\fett{p}_3,\fett{p}_2,\fett{p}_1) 
= {}_{{\rm \MakeUppercase{ \romannumeral 1}} 
 \oplus {\rm \MakeUppercase{ \romannumeral 2}} }C_{jik}^{(213)} (\fett{p}_2,\fett{p}_1,\fett{p}_3)\,,
\end{align}  
  \begin{align}
& C_{ijkl}^{(1122)} (\fett{p}_1,\fett{p}_2,\fett{p}_3, \fett{p}_4) 
 = \,C_{jkli}^{(1221)} (\fett{p}_2,\fett{p}_3,\fett{p}_4, \fett{p}_1) = 
  C_{klij}^{(2211)} (\fett{p}_3,\fett{p}_4,\fett{p}_1, \fett{p}_2) \nonumber \\
& \hspace{10mm} =   C_{lijk}^{(2112)} (\fett{p}_4,\fett{p}_1,\fett{p}_2, \fett{p}_3)    = \,C_{likj}^{(2121)} (\fett{p}_4,\fett{p}_1,\fett{p}_3, \fett{p}_2) 
=  C_{ikjl}^{(1212)} (\fett{p}_1,\fett{p}_3,\fett{p}_2, \fett{p}_4) \nonumber \\
&   \hspace{10mm} = -4 \, S_i^{(1)} (\fett{p}_1) \, P_L(p_1)\, S_j^{(1)}(\fett{p}_2)\, P_L(p_2)  \\
    & \ws \hspace{10mm} \times  \left[ S_k^{(2)}(-\fett{p}_1, \fett{p}_{13} ) 
  S_l^{(2)}(\fett{p}_2, \fett{p}_{13} ) P_L (p_{13}) +  
 S_k^{(2)}(-\fett{p}_2, \fett{p}_{23} ) S_l^{(2)}(\fett{p}_1, \fett{p}_{23} ) P_L (p_{23})    \right], \nonumber \\
&C_{ijkl}^{(1113)} (\fett{p}_1,\fett{p}_2,\fett{p}_3, \fett{p}_4)  = \, C_{jkli}^{(1131)} (\fett{p}_2,\fett{p}_3,\fett{p}_4, \fett{p}_1) \nonumber \\
&\hspace{10mm} = C_{klij}^{(1311)} (\fett{p}_3,\fett{p}_4,\fett{p}_1, \fett{p}_2) 
  = C_{lijk}^{(3111)} (\fett{p}_4,\fett{p}_1,\fett{p}_2, \fett{p}_3) \nonumber \\
  & \hspace{10mm} =\,-6 \,S_i^{(1)}(\fett{p}_1) \,S_j^{(1)}(\fett{p}_2) \,
  S_k^{(1)}(\fett{p}_3) \label{c3111}
  \,S_l^{(3)}(\fett{p}_1,\fett{p}_2,\fett{p}_3) \,P_L(p_1)\,P_L(p_2)\,P_L(p_3)\, .
\end{align}

\subsection{Non-Gaussian initial conditions\label{app3.2}}

As shown in equation~(\ref{eq:nongauss}), additional terms arise in case the initial conditions contain non-Gaussian contributions.
\begin{align}
&C_{ij}^{(12)} (\fett{p}) = C_{ji}^{(21)} (\fett{p})
 =  \int \frac{\dd^3 p'}{(2\pi)^3} S_i^{(1)}(\fett{p})  \, S_j^{(2)}(\fett{p}', \fett{p}-\fett{p}') \,B_0 (p,p', |p-p'|) \,,  \\
&C_{ijk}^{(111)} (\fett{p}_1,\fett{p}_2,\fett{p}_3) =  
   \ii\, S_i^{(1)}(\fett{p}_1)\,S_j^{(1)}(\fett{p}_2)\,S_k^{(1)}(\fett{p}_3)\,B_0(p_1,p_2,p_3) \,,   \\
  &{}_{{\rm \MakeUppercase{ \romannumeral 2}}}C_{ijk}^{(112)}(\fett{p}_1,\fett{p}_2,\fett{p}_3) 
  = {}_{{\rm \MakeUppercase{ \romannumeral 2}}}C_{jki}^{(121)}
 (\fett{p}_2,\fett{p}_3,\fett{p}_1) = {}_{{\rm \MakeUppercase{ \romannumeral 2}}}C_{kij}^{(211)}
 (\fett{p}_3,\fett{p}_1,\fett{p}_2) \,, \nonumber  \\
  &\hspace{10mm} = \,\ii\,S_i^{(1)}(\fett{p}_1)\,S_j^{(1)}(\fett{p}_2) 
  \int \frac{\dd^3 p'}{(2\pi)^3} S_k^{(2)}(\fett{p}',\fett{p}_3-\fett{p}')\,  %
  T_0(\fett{p}_1, \fett{p}_2, \fett{p}',\fett{p}_3-\fett{p}')\, ,  \\ 
  &{}_{{\rm \MakeUppercase{ \romannumeral 1}}}C_{ijk}^{(113)}(\fett{p}_1,\fett{p}_2,\fett{p}_3) 
 = \, 3\ii\,B_0(\fett{p}_1,\fett{p}_2, \fett{p}_3)\,\nonumber \\
  & \hspace{4.cm} \times \int \frac{\dd^3 p'}{(2\pi)^3} S_i^{(1)}(\fett{p}_1)
 \,S_j^{(1)}(\fett{p}_2)\,S_k^{(3)}(\fett{p}_3,\fett{p}',-\fett{p}')\,P_L(p') \,, \\ %
 &{}_{{\rm \MakeUppercase{ \romannumeral 2}}}C_{ijk}^{(113)}(\fett{p}_1,\fett{p}_2,\fett{p}_3) = 
 \, -3\ii \,P_L(p_1) \,\int \frac{\dd^3 p'}{(2\pi)^3} S_i^{(1)}(\fett{p}_1)\,S_j^{(1)}(\fett{p}_2)
 \nonumber \\
   &\hspace{2.5cm} \times S_k^{(3)}(\fett{p}_1, \fett{p}', \fett{p}_2-\fett{p}')\,B_0(p_2,p',|p_2-p'|)
 - \left( \fett{p}_1  \leftrightarrow \fett{p}_2, \, \ii \leftrightarrow j \right) \,, \\ 
 &{}_{{\rm \MakeUppercase{ \romannumeral 1}} \oplus {\rm \MakeUppercase{ \romannumeral 2}} }C_{ijk}^{(113)}
   (\fett{p}_1,\fett{p}_2,\fett{p}_3)  
\equiv    {}_{\rm \MakeUppercase{ \romannumeral 1}}C_{ijk}^{(113)}
   (\fett{p}_1,\fett{p}_2,\fett{p}_3)  +   {}_{\rm \MakeUppercase{ \romannumeral 2}}C_{ijk}^{(113)}
   (\fett{p}_1,\fett{p}_2,\fett{p}_3)  \nonumber \\
  & \hspace{35mm}     = {}_{{\rm \MakeUppercase{ \romannumeral 1}} \oplus {\rm \MakeUppercase{ \romannumeral 2}} }C_{jki}^{(131)}
  (\fett{p}_2,\fett{p}_3,\fett{p}_1) 
         = {}_{{\rm \MakeUppercase{ \romannumeral 1}}
 \oplus {\rm \MakeUppercase{ \romannumeral 2}} }C_{kij}^{(311)}(\fett{p}_3,\fett{p}_1,\fett{p}_2) 
   \,, \\
  &{}_{{\rm \MakeUppercase{ \romannumeral 1}}}C_{ijk}^{(122)}(\fett{p}_1,\fett{p}_2,\fett{p}_3)
  = \,-2\ii\,S_i^{(1)}(\fett{p}_1)\,P_L(p_1) \,S_j^{(2)}(\fett{p}_1,\fett{p}_3) \nonumber \\
 &\hspace{15mm} \times \int \frac{\dd^3 p'}{(2\pi)^3} S_k^{(2)}(\fett{p}',\fett{p}_3-\fett{p}')
   \,B_0\left(p',p_3,|p'-p_3|\right)
-  \left( \fett{p}_3  \leftrightarrow \fett{p}_2, j \leftrightarrow k \right) \,, \\  %
   &{}_{{\rm \MakeUppercase{ \romannumeral 2}}}C_{ijk}^{(122)}(\fett{p}_1,\fett{p}_2,\fett{p}_3) =
 \, -4\ii \,S_i^{(1)}(\fett{p}_1) \,\int \frac{\dd^3 p'}{(2\pi)^3} S_j^{(2)}(\fett{p}',\fett{p}_2-\fett{p}')  %
  \nonumber \\  
    &\hspace{35mm}\times \,S_k^{(2)}(\fett{p}_1+\fett{p}', \fett{p}_2
 -\fett{p}')\,P_L(|p_2-p'|)\,B_0(p_1,p', |p_1+p'|) \,, \\
 &{}_{{\rm \MakeUppercase{ \romannumeral 1}} \oplus {\rm \MakeUppercase{ \romannumeral 2}} }C_{ijk}^{(122)}
 (\fett{p}_1,\fett{p}_2,\fett{p}_3)  
 \equiv {}_{\rm \MakeUppercase{ \romannumeral 1}}C_{ijk}^{(122)}
 (\fett{p}_1,\fett{p}_2,\fett{p}_3)  +  {}_{\rm \MakeUppercase{ \romannumeral 2}}C_{ijk}^{(122)}
 (\fett{p}_1,\fett{p}_2,\fett{p}_3) \nonumber \\
&\hspace{35mm}  = {}_{{\rm \MakeUppercase{ \romannumeral 1}}
  \oplus {\rm \MakeUppercase{ \romannumeral 2}} }C_{jki}^{(221)}(\fett{p}_2,\fett{p}_3,\fett{p}_1) 
         = {}_{{\rm \MakeUppercase{ \romannumeral 1}}
  \oplus {\rm \MakeUppercase{ \romannumeral 2}} }C_{kij}^{(212)}(\fett{p}_3,\fett{p}_1,\fett{p}_2)  \,, 
\end{align}
\begin{align}  
  C_{ijkl}^{(1111)} &(\fett{p}_1,\fett{p}_2,\fett{p}_3,\fett{p}_4) =
    S_i^{(1)}(\fett{p}_1)\,S_j^{(1)}(\fett{p}_2)\,S_k^{(1)}(\fett{p}_3)\,S_k^{(1)}(\fett{p}_3)\,S_l^{(1)}(\fett{p}_4)  \nonumber \\
   &  \hspace{5cm}\times 
   T_0(\fett{p}_1,\fett{p}_2,\fett{p}_3,\fett{p}_4) \,, \\
C _{ijkl}^{(1112)}&(\fett{p}_1,\fett{p}_2,\fett{p}_3,\fett{p}_4) 
= C _{jkli}^{(1121)}(\fett{p}_2,\fett{p}_3,\fett{p}_4,\fett{p}_1)  \nonumber \\
   &=  C _{klij}^{(1211)}(\fett{p}_3,\fett{p}_4,\fett{p}_1,\fett{p}_2) =
     C _{lijk}^{(2111)}(\fett{p}_4,\fett{p}_1,\fett{p}_2,\fett{p}_3) \, \nonumber \\
 &= -2\,S_i^{(1)}(\fett{p}_1) P_L(p_1)\, S_j^{(1)}(\fett{p}_2)\,
 S_k^{(1)}(\fett{p}_3)\,S_l^{(2)}(\fett{p}_1,\fett{p}_{23})
  \,B_0(p_2,p_3,p_{23}) \nonumber \\
  &\ws\hspace{0.33cm}-2\,S_i^{(1)}(\fett{p}_1) \,
  S_j^{(1)}(\fett{p}_2)\,P_L(p_2) S_k^{(1)}(\fett{p}_3)\,S_l^{(2)}(\fett{p}_2,\fett{p}_{13})
  \,B_0(p_1,p_3,p_{13}) \nonumber \\
  &\ws\hspace{0.33cm}-2\,S_i^{(1)}(\fett{p}_1) \, S_j^{(1)}(\fett{p}_2)\,
 S_k^{(1)}(\fett{p}_3)\,P_L(p_3)S_l^{(2)}(\fett{p}_3,\fett{p}_{12})
  \,B_0(p_1,p_2,p_{12})  \,.
\end{align}

\section{From primordial curvature perturbations to non-Gaussian initial conditions\label{app2}}

Primordial non-Gaussianity from inflation is usually quantified in terms of the $N$-point statistics of the superhorizon metric perturbations $\Phi(\fett{x})$, extrapolated to 
the epoch of matter domination.
For our bispectrum considerations in section~\ref{sec:realspace}, up to one-loop order only the primordial bispectrum $B_\Phi (k_1,k_2,k_3)$
and the primordial trispectrum $T_\Phi(\fett{k}_1, \fett{k}_2, \fett{k}_3, \fett{k}_4)$ play a role.
To make use of these initial conditions, however, we must first link  $B_\Phi$ and $T_\Phi$ respectively to 
the linear matter bispectrum $B_0$  and trispectrum $T_0$  at redshift $z_0$ via
\begin{align}
 &B_0 (k_1,k_2,k_3) = M(k_1,z_0)\, M(k_2,z_0)\, M(k_3,z_0)\,B_\Phi (k_1,k_2,k_3) \,, \nonumber  \\
 &T_0(\fett{k}_1, \fett{k}_2, \fett{k}_3, \fett{k}_4) = M(k_1,z_0)\, M(k_2,z_0)\, M(k_3,z_0) \,M(k_4,z_0) \, 
  T_\Phi(\fett{k}_1, \fett{k}_2, \fett{k}_3, \fett{k}_4)  \,, \\
\intertext{where}
  &M(k,z)  = \frac 2 3 \frac{k^2 T(k)D(z)}{\Omega_m H_0^2}\,.
\end{align}
Here, the linear transfer function $T(k)$ asymptotes to unity as $k \to0$, while
the linear growth function $D(z)$ is defined to coincide with the scale factor $a$ during matter domination.  Note that this definition of $D(z)$ differs from 
our previous definition in sections~\ref{sec:formalism} and~\ref{sec:realspace}.
We compute $M(k,z)$ in this work using the publicly available Boltzmann code {\texttt{CAMB}}~\cite{Lewis:1999bs}. 

Assuming primordial non-Gaussianity of the local type, one can further parameterise the degree of PNG in 
terms of the parameter $f_{\rm NL}$, defined  via~\cite{Bartolo:2004if}
\begin{align}
  \Phi(\fett{x}) = \varphi(\fett{x}) +f_{\rm NL} \left[ \varphi^2(\fett{x}) 
 - \langle \varphi^2(\fett{x}) \rangle  \right]\,,
\end{align}
where $\varphi(\fett{x})$ is a Gaussian field.
The primordial bispectrum and trispectrum of $\Phi$ are then given by
\begin{align}
  &B_\Phi (k_1,k_2,k_3) = 2 f_{{\rm{NL}}}\,P_{\Phi}({k}_1) \,P_{\Phi}({k}_2) + {\rm{two\,\, perms.}} \,, \nonumber \\
  &T_\Phi(\fett{k}_1, \fett{k}_2, \fett{k}_3, \fett{k}_4) = 
 4 f_{{\rm NL}}^2 \,P_{\Phi}({k}_1) \,P_{\Phi}({k}_2) \left[ P_{\Phi}({k}_{13})+P_{\Phi}({k}_{14})  \right] 
 + 5 \,{\rm{perms.}} \,,
\end{align}
to leading order in $f_{\rm NL}$.

\section{One-loop expressions for the bispectrum\label{app4}}

\subsection{Gaussian initial conditions\label{app4.1}}

We split up the one-loop contribution to the matter bispectrum into five groups:
\be
B_{{\rm Gaussian}}^{(1)} =  B_{411 \oplus 123 \oplus 222} + B_{1122 \oplus 1113}
 + B_{{\rm{xx}} \otimes {\rm{yy}}} +B_{11 \otimes 211}  - B_{11 \otimes 11 \otimes 11} \,,
\ee 
where
\be
  B_{411 \oplus 123 \oplus 222}    =\,-\ii k_{1i}k_{2j}k_{3k} \,C_{ijk}^{\left\{(411\oplus123 \oplus 222)\right\}}(\fett{k}_1,\fett{k}_2,\fett{k}_3)  \,, 
\ee  
\begin{align}  
 B_{1122 \oplus 1113}  &=\, \frac 1 2 k_{1i}k_{1j}k_{2k}k_{3l} \int \frac{\dd^3 p}{(2\pi)^3} \, C_{ijkl}^{\left\{(1122 \oplus 1113)\right\}} (\fett{p}, \fett{k}_1 - \fett{p}, \fett{k}_2, \fett{k}_3) \nonumber \\
 &\ws+ \frac 1 2  k_{2i}k_{2j}k_{1k}k_{3l} \int \frac{\dd^3 p}{(2\pi)^3} \, C_{ijkl}^{\left\{(1122 \oplus 1113)\right\}} (\fett{p}, \fett{k}_2 - \fett{p}, \fett{k}_1, \fett{k}_3) \nonumber \\
 &\ws+ \frac 1 2  k_{3i}k_{3j}k_{1k}k_{2l} \int \frac{\dd^3 p}{(2\pi)^3} \, C_{ijkl}^{\left\{(1122 \oplus 1113)\right\}} (\fett{p}, \fett{k}_3 - \fett{p}, \fett{k}_1, \fett{k}_2)  \,, 
\end{align} 
 \begin{align}
 B_{{\rm{xx}} \otimes {\rm{yy}}} &=\,  k_{1i} k_{2j}k_{2l} k_{3m} \left[ C_{ij}^{(11)}(\fett{k}_1)\, C_{lm}^{\left\{(31 \oplus 22)\right\}}(\fett{k}_3)
  + C_{ij}^{\left\{(31 \oplus 22)\right\}}(\fett{k}_1) \,C_{lm}^{(11)}(\fett{k}_3)   \right] \nonumber \\
 &\ws+ k_{1i} k_{2j} k_{1l} k_{3m} \left[ C_{ij}^{(11)}(\fett{k}_2) \,  C_{lm}^{\left\{(31 \oplus 22)\right\}}(\fett{k}_3)
  +C_{ij}^{\left\{(31 \oplus 22)\right\}}(\fett{k}_2)\,C_{lm}^{(11)}(\fett{k}_3)   \right] \nonumber \\
&\ws+ k_{1i} k_{3j} k_{2l} k_{3m} \left[ C_{ij}^{(11)}(\fett{k}_1) \, C_{lm}^{\left\{(31 \oplus 22)\right\}}(\fett{k}_2)
  +  C_{ij}^{\left\{(31 \oplus 22)\right\}}(\fett{k}_1) \,C_{lm}^{(11)}(\fett{k}_2)   \right] \,, 
\end{align}
\begin{align} 
&B_{11 \otimes 211} = \,\frac{\ii}{2} k_{1i}k_{2j}k_{3k}\,C_{ijk}^{\left\{(211)\right\}}(\fett{k}_1, \fett{k}_2,\fett{k}_3)  \nonumber\\
 &\qquad  \times \intp \left[ k_{1l} k_{1m}+k_{2l} k_{2m}+k_{3l} k_{3m}\right]\, C_{lm}^{(11)}(\fett{p}) \nonumber \\
 &\qquad+\ii \intp k_{1i} k_{2j}\, C_{ij}^{(11)}(\fett{p})\, k_{1k}k_{2l}k_{3m} \,C_{klm}^{\left\{(211)\right\}} (\fett{k}_1 -\fett{p}, \fett{p}+\fett{k}_2, \fett{k}_3) \nonumber \\
&\qquad+\ii \intp k_{1i} k_{3j}\, C_{ij}^{(11)}(\fett{p})\, k_{1k}k_{2l}k_{3m} \,C_{klm}^{\left\{(211)\right\}} (\fett{k}_1 -\fett{p}, \fett{k}_2,\fett{p}+ \fett{k}_3) \nonumber \\
&\qquad+\ii \intp k_{2i} k_{3j}\, C_{ij}^{(11)}(\fett{p})\, k_{1k}k_{2l}k_{3m} \,C_{klm}^{\left\{(211)\right\}} (\fett{k}_1, \fett{k}_2-\fett{p},\fett{p}+ \fett{k}_3)\\
&\qquad+ \frac \ii 2 k_{1i}k_{2j}\,C_{ij}^{(11)}(\fett{k}_1) \intp \left[  k_{3k}k_{3l}k_{2m} -k_{2k}k_{2l}k_{3m}   \right]\,C_{klm}^{\left\{(211)\right\}}(\fett{p}, \fett{k}_3 -\fett{p}, -\fett{k}_3)
 \nonumber \\
&\qquad+\frac \ii 2 k_{1i}k_{2j}\,C_{ij}^{(11)}(\fett{k}_2) \intp \left[ k_{3k}k_{3l}k_{1m} - k_{1k}k_{1l}k_{3m}  \right] \,C_{klm}^{\left\{(211)\right\}}(\fett{p},\fett{k}_3-\fett{p},-\fett{k}_3)    \nonumber \\
&\qquad+ \frac \ii 2 k_{1i}k_{3j}\,C_{ij}^{(11)}(\fett{k}_1) \intp \left[ k_{2k}k_{2l}k_{3m} - k_{3k}k_{3l}k_{2m} \right]\,C_{klm}^{\left\{(211)\right\}}(\fett{p},\fett{k}_2-\fett{p},-\fett{k}_2)  \nonumber \\
&\qquad+ \frac \ii 2 k_{1i}k_{3j}\,C_{ij}^{(11)}(\fett{k}_3) \intp \left[ k_{2k}k_{2l}k_{1m} - k_{1k}k_{1l}k_{2m}\right] \,C_{klm}^{\left\{(211)\right\}}(\fett{p},\fett{k}_2-\fett{p},-\fett{k}_2) \nonumber \\
&\qquad+\frac \ii 2 k_{2i}k_{3j}\,C_{ij}^{(11)}(\fett{k}_2) \intp \left[ k_{1k}k_{1l}k_{3m} - k_{3k}k_{3l}k_{1m} \right]\,C_{klm}^{\left\{(211)\right\}}(\fett{p},\fett{k}_1-\fett{p},-\fett{k}_1) \nonumber \\
&\qquad+\frac \ii 2 k_{2i}k_{3j}\,C_{ij}^{(11)}(\fett{k}_3) \intp \left[ k_{1k}k_{1l}k_{2m} - k_{2k}k_{2l}k_{1m}  \right]\,C_{klm}^{\left\{(211)\right\}}(\fett{p},\fett{k}_1-\fett{p},-\fett{k}_1) \,, \nonumber 
\end{align}
\begin{align}
 &B_{11 \otimes 11 \otimes 11} = \, \frac{1}{2}  
  \left(k_{1m}k_{1n}+k_{2m}k_{2n}+k_{3m}k_{3n} \right) \int \frac{\dd^3 p}{(2\pi)^3} C_{mn}^{(11)}(\fett{p}) \,  \nonumber \\ 
 &\ws \hspace{0.5cm}  \times \Big[  k_{1i}k_{2j} C_{ij}^{(11)}(\fett{k}_2)\,k_{1k}k_{3l} C_{kl}^{(11)}(\fett{k}_3)   + k_{1i}k_{2j}\, C_{ij}^{(11)}(\fett{k}_1)\,k_{2k}k_{3l} C_{kl}^{(11)}(\fett{k}_3) \Big.  \nonumber \\ 
 &\ws \hspace{2.6cm} \Big.
 +  k_{1i}k_{3j}\, C_{ij}^{(11)}(\fett{k}_1)\,k_{2k}k_{3l} C_{kl}^{(11)}(\fett{k}_2)  \Big] \nonumber \\
 &\ws \hspace{0.5cm}+\frac{1}{2} k_{1i}k_{3j}\, C_{ij}^{(11)}(\fett{k}_3) \int \frac{\dd^3 p}{(2\pi)^3} k_{1k}k_{2l} \,C_{kl}^{(11)}(\fett{p}) \, k_{1m}k_{2n} \, C_{mn}^{(11)}(\fett{p}+\fett{k}_2) \nonumber \\
 &\ws \hspace{0.5cm}+\frac{1}{2} k_{2i}k_{3j}\, C_{ij}^{(11)}(\fett{k}_3) \int \frac{\dd^3 p}{(2\pi)^3} k_{1k}k_{2l} \,C_{kl}^{(11)}(\fett{p}) \, k_{1m}k_{2n} \, C_{mn}^{(11)}(\fett{p}+\fett{k}_1) \nonumber \\
 &\ws \hspace{0.5cm}+\frac{1}{2} k_{1i}k_{2j}\, C_{ij}^{(11)}(\fett{k}_2) \int \frac{\dd^3 p}{(2\pi)^3} k_{1k}k_{3l} \,C_{kl}^{(11)}(\fett{p}) \, k_{1m}k_{3n} \, C_{mn}^{(11)}(\fett{p}+\fett{k}_3)   \\
 &\ws \hspace{0.5cm}+\frac{1}{2} k_{1i}k_{2j}\, C_{ij}^{(11)}(\fett{k}_1) \int \frac{\dd^3 p}{(2\pi)^3} k_{2k}k_{3l} \,C_{kl}^{(11)}(\fett{p}) \, k_{2m}k_{3n} \, C_{mn}^{(11)}(\fett{p}+\fett{k}_3) \nonumber \\
 &\ws \hspace{0.5cm}+\frac{1}{2} k_{2i}k_{3j}\, C_{ij}^{(11)}(\fett{k}_2) \int \frac{\dd^3 p}{(2\pi)^3} k_{1k}k_{3l} \,C_{kl}^{(11)}(\fett{p}) \, k_{1m}k_{3n} \, C_{mn}^{(11)}(\fett{p}+\fett{k}_1) \nonumber \\
 &\ws \hspace{0.5cm}+\frac{1}{2} k_{1i}k_{3j}\, C_{ij}^{(11)}(\fett{k}_1) \int \frac{\dd^3 p}{(2\pi)^3} k_{2k}k_{3l} \,C_{kl}^{(11)}(\fett{p}) \, k_{2m}k_{3n} \, C_{mn}^{(11)}(\fett{p}+\fett{k}_2) \nonumber \\
 &\ws \hspace{0.5cm}+\int \frac{\dd^3 p}{(2\pi)^3} k_{1i}k_{2j}\, C_{ij}^{(11)}(\fett{p}) \, k_{1k}k_{3l} \,C_{kl}^{(11)}(\fett{p}-\fett{k}_1) \, k_{2m}k_{3n} \, C_{mn}^{(11)}(\fett{p}+\fett{k}_2)  \,.  \nonumber 
\end{align}
Note that we have employed a shorthand notation, e.g., $C_{ijkl}^{\left\{(1122 \oplus 1113)\right\}} \equiv C_{ijkl}^{\left\{(1122)\right\}} + C_{ijkl}^{\left\{(1113)\right\}}$,
where $\{\cdots\}$ indicates summation over all possible correlator settings, e.g., $C_{ijkl}^{\left\{(1113)\right\}}= C_{ijkl}^{(1113)} + C_{ijkl}^{(1131)}+ C_{ijkl}^{(1311)}+C_{ijkl}^{(3111)}$,   
and $C_{ijk}^{\left\{(123) \right\}} $ is understood to mean 
${}_{{\rm \MakeUppercase{ \romannumeral 1}} \oplus \rm \MakeUppercase{ \romannumeral 2}} C_{ijk}^{\left\{(123)\right\}}$ .
As usual, summation over repeated indices $\in \{i,j,k,l\}$ is implied.

\subsection{Non-Gaussian initial conditions\label{app4.2}}

The one-loop terms arising from non-Gaussian initial conditions are
\be 
  B_{112 \oplus 122 \oplus 113} = \,-\ii k_{1i}k_{2j}k_{3k} C_{ijk}^{\left\{(112 \oplus 122 \oplus 113)\right\}}(\fett{k}_1,\fett{k}_2,\fett{k}_3) \,, 
\ee  
\begin{align}
 B_{11\otimes12} &=\, k_{1i} k_{2j}k_{2l} k_{3m} \left[ C_{ij}^{(11)}(\fett{k}_1)\, C_{lm}^{\left\{(12)\right\}}(\fett{k}_3)
  + C_{ij}^{\left\{(12)\right\}}(\fett{k}_1) \,C_{lm}^{(11)}(\fett{k}_3)   \right] \nonumber \\
 &\ws+ k_{1i} k_{2j} k_{1l} k_{3m} \left[ C_{ij}^{(11)}(\fett{k}_2) \,  C_{lm}^{\left\{(12)\right\}}(\fett{k}_3)
  +C_{ij}^{\left\{(12)\right\}}(\fett{k}_2)\,C_{lm}^{(11)}(\fett{k}_3)   \right] \nonumber \\
 &\ws+ k_{1i} k_{3j} k_{2l} k_{3m} \left[ C_{ij}^{(11)}(\fett{k}_1) \, C_{lm}^{\left\{(12)\right\}}(\fett{k}_2)
  +  C_{ij}^{\left\{(12)\right\}}(\fett{k}_1) \,C_{lm}^{(11)}(\fett{k}_2)   \right] \,, 
\end{align}
\begin{align}
B_{1111 \oplus 1112}   &=\, \frac{1}{2}k_{1i}k_{1j}k_{2k}k_{3l} \int \frac{\dd^3 p}{(2\pi)^3} 
   \, C_{ijkl}^{\left\{(1111 \oplus 1112)\right\}} (\fett{p}, \fett{k}_1 - \fett{p}, \fett{k}_2, \fett{k}_3) \nonumber \\
 &\ws+ \frac{1}{2}k_{2i}k_{2j}k_{1k}k_{3l} \int \frac{\dd^3 p}{(2\pi)^3} \, C_{ijkl}^{\left\{(1111 \oplus 1112)\right\}} (\fett{p}, \fett{k}_2 - \fett{p}, \fett{k}_1, \fett{k}_3) \nonumber  \\
 &\ws+ \frac{1}{2}k_{3i}k_{3j}k_{1k}k_{2l} \int \frac{\dd^3 p}{(2\pi)^3} \, C_{ijkl}^{\left\{(1111 \oplus 1112)\right\}} (\fett{p}, \fett{k}_3 - \fett{p}, \fett{k}_1, \fett{k}_2)    \, ,
\end{align}
\begin{align}
 &B_{11 \otimes 111} = \,   B_{11 \otimes 211} (  11\otimes 211  \rightarrow 11 \otimes 111)   \nonumber \\ 
 &\ws \hspace{0.5cm}= \, \frac{\ii}{2} k_{1i}k_{2j}k_{3k}\,C_{ijk}^{(111)}(\fett{k}_1, \fett{k}_2,\fett{k}_3)  
                                       \intp \left[ k_{1l} k_{1m}+k_{2l} k_{2m}+k_{3l} k_{3m}\right]\, C_{lm}^{(11)}(\fett{p}) \nonumber \\
 &\ws \hspace{1,cm}+ \ii \intp k_{1i} k_{2j}\, C_{ij}^{(11)}(\fett{p})\, k_{1k}k_{2l}k_{3m} \,C_{klm}^{(111)} (\fett{k}_1 -\fett{p}, \fett{p}+\fett{k}_2, \fett{k}_3) \nonumber \\
&\ws \hspace{1cm}+\ii \intp k_{1i} k_{3j}\, C_{ij}^{(11)}(\fett{p})\, k_{1k}k_{2l}k_{3m} \,C_{klm}^{(111)} (\fett{k}_1 -\fett{p}, \fett{k}_2,\fett{p}+ \fett{k}_3) \nonumber \\
&\ws \hspace{1cm}+ \ii \intp k_{2i} k_{3j}\, C_{ij}^{(11)}(\fett{p})\, k_{1k}k_{2l}k_{3m} \,C_{klm}^{(111)} (\fett{k}_1, \fett{k}_2-\fett{p},\fett{p}+ \fett{k}_3) \\
&\ws \hspace{1cm}+ \frac \ii 2 k_{1i}k_{2j}\,C_{ij}^{(11)}(\fett{k}_1) \intp \left[  k_{3k}k_{3l}k_{2m} -k_{2k}k_{2l}k_{3m}   \right]\,C_{klm}^{(111)}(\fett{p}, \fett{k}_3 -\fett{p}, -\fett{k}_3)
 \nonumber \\
&\ws \hspace{1cm}+ \frac \ii 2 k_{1i}k_{2j}\,C_{ij}^{(11)}(\fett{k}_2) \intp \left[ k_{3k}k_{3l}k_{1m} - k_{1k}k_{1l}k_{3m}  \right] \,C_{klm}^{(111)}(\fett{p},\fett{k}_3-\fett{p},-\fett{k}_3)    \nonumber \\
&\ws \hspace{1cm}+ \frac \ii 2 k_{1i}k_{3j}\,C_{ij}^{(11)}(\fett{k}_1) \intp \left[ k_{2k}k_{2l}k_{3m} - k_{3k}k_{3l}k_{2m} \right]\,C_{klm}^{(111)}(\fett{p},\fett{k}_2-\fett{p},-\fett{k}_2)  \nonumber \\
&\ws \hspace{1cm}+ \frac \ii 2 k_{1i}k_{3j}\,C_{ij}^{(11)}(\fett{k}_3) \intp \left[ k_{2k}k_{2l}k_{1m} - k_{1k}k_{1l}k_{2m}\right] \,C_{klm}^{(111)}(\fett{p},\fett{k}_2-\fett{p},-\fett{k}_2) \nonumber \\
&\ws \hspace{1cm}+ \frac \ii 2 k_{2i}k_{3j}\,C_{ij}^{(11)}(\fett{k}_2) \intp \left[ k_{1k}k_{1l}k_{3m} - k_{3k}k_{3l}k_{1m} \right]\,C_{klm}^{(111)}(\fett{p},\fett{k}_1-\fett{p},-\fett{k}_1) \nonumber \\
&\ws \hspace{1cm}+ \frac \ii 2 k_{2i}k_{3j}\,C_{ij}^{(11)}(\fett{k}_3) \intp \left[ k_{1k}k_{1l}k_{2m} - k_{2k}k_{2l}k_{1m}  \right]\,C_{klm}^{(111)}(\fett{p},\fett{k}_1-\fett{p},-\fett{k}_1) \, , \nonumber 
\end{align}
again with the understanding that $C_{ijk}^{\left\{ (122)\right\}} \equiv {}_{{\rm \MakeUppercase{ \romannumeral 1}} \oplus \rm \MakeUppercase{ \romannumeral 2}} C_{ijk}^{\left\{(122)\right\}}$,  and so forth.

\section{Diagrams and the relation between the SPT and the LPT bispectra\label{app5}}

In this section we demonstrate how to construct diagrams in LPT.  Consider the diagram in figure~\ref{fig:explain_diagrams}, corresponding to the contribution:
\begin{align}\label{eq:321I}
  &-\frac{i}{2}  k_{1i} k_{2j} C_{ij}^{(11)}(\fett{k}_1) k_{3k} k_{3l} k_{2m}   
   C_{klm}^{(112)}(\fett{p}, \fett{k}_3-\fett{p},- \fett{k}_3) \stackrel{\rm eq.\,(\ref{c211})}{=}  \nonumber \\&
  \hspace{18mm}   - k_{1i} S_{i}^{(1)}\,(\fett{k}_1)P_L(k_1)\,k_{2j} S_{j}^{(1)}(\fett{k}_1) \nonumber \\&
   \hspace{22mm} \times 
   k_{3k} S_{k}^{(1)}(\fett{p}) P_L(p) k_{3l} S_{l}^{(1)}(\fett{k}_3-\fett{p})  P_L(|\fett{k}_3-\fett{p}|)  
   k_{2m} S_{m}^{(2)}(\fett{p}, \fett{k}_3-\fett{p}) \,,
\end{align}
where for simplicity we have omitted the integration over $\fett{p$}.  
The RHS of equation~(\ref{eq:321I}) can be split into five parts:
(i) $k_{1i} S_i^{(1)}(\fett{k}_1) P^{1/2}_L(k_1)$,
(ii) $k_{2j} S_j^{(1)}(\fett{k}_1) P^{1/2}_L(k_1)$,
(iii) $k_{3k} S_k^{(1)}(\fett{p}) P^{1/2}_L(p)$,
(iv)  $k_{3l} S_l^{(1)}(\fett{k}_3-\fett{p}) P^{1/2}_L(|\fett{k}_3-\fett{p}|)$, and 
(v) $k_{2m} S_{m}^{(2)}(\fett{p}, \fett{k}_3-\fett{p})  P^{1/2}_L(p) P^{1/2}_L(|\fett{k}_3-\fett{p}|)$.

\begin{figure}[t]
\hspace*{1.5cm}
\begin{centering}
\hskip3cm\includegraphics[width=0.44\textwidth]{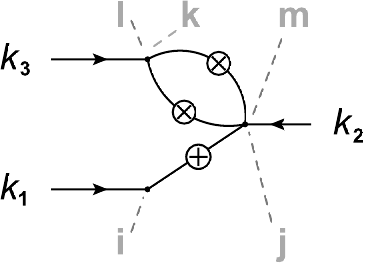}
\end{centering}
\caption{An example diagram contributing to the one-loop bispectrum.   The letters $\{i,j,k,l,m\}$ are indices of the correlators $C_{ijk\cdots}$, while $\fett{k}_1+\fett{k}_2+\fett{k}_3 = \fett{0}$ are the wavevectors defining the triangle of  the bispectrum.  The symbol $\oplus$ denotes an initial  linear power spectrum.
\label{fig:explain_diagrams}}
\end{figure}

\mbox{Part~(i)}
corresponds to an external vector $\fett{k}_1$ entering at the point~$i$, from whence emerges an internal line of momentum $\fett{k}_1$ (represented by  $S_i^{(1)}(\fett{k}_1)$),
terminating at the linear power spectrum $P_L(k_1)$.  Parts~(ii) to~(iv) bear similar interpretations.    Part~(v) has an external line $\fett{k}_2$ entering at the point~$m$, and two internal lines with momenta $\fett{p}$ and $\fett{k}_3-\fett{p}$ emanating from it (indicated by $S_{m}^{(2)}(\fett{p}, \fett{k}_3-\fett{p})$), which terminate respectively at the power spectra  $P_L(p)$ and  $P_L(|\fett{k}_3-\fett{p}|)$.    Parts~(i) and (ii) are joined together at their common linear power spectrum $P_L(k_1)$, thereby linking the point~$i$ to the point~$j$.  Likewise,
parts~(iii) is joined to one of the internal lines of part~(v) at their common power spectrum $P_L(p)$, thus connecting the points~$m$ and~$k$.  The joining of parts~(iv) and~(v) at the power 
spectrum $P_L(|\fett{k}_3-\fett{p})$ results in a connection between the points $m$ and $l$.  Finally, the points $k$ and $l$ are fused into one single point because of the common external vector~$\fett{k}_3$ in parts~(iii) and~(iv), in the same manner that $j$ and $m$ are fused together because of the common external vector~$\fett{k}_2$ appearing in parts~(ii) and~(v).  
 The resulting diagram contains a linear power spectrum in every internal line, and 
 belongs to the class of ${}_{\rm I}B_{321}$: three internal lines emerge from~$\{j,m\}$, two from the point~$\{k,l\}$, and one from the point~$i$ (see also diagram (b) in figure~\ref{fig:diagrams}).

The construction of all other diagrams is then completely straightforward.   Only one new rule needs to be introduced in the case of primordial non-Gaussianity: while the linear power spectrum has only one ingoing and one outgoing momentum, the initial bispectrum (trispectrum) has three (four) entries/exits.

One advantage of using these quasi-Feynman rules is that it is  now possible to associate all LPT contributions in equation~(\ref{oneloop}) to their counterparts in SPT,
since both approaches should lead to the same diagrams for the $N$-point  function.   Note, however, that for every SPT diagram there are several LPT contributions.  The number of LPT  contributions to each diagram is listed in tables~\ref{tab:diagramsGaussian} and~\ref{tab:nongauss}.

\begin{table}[t]
\caption{The number of LPT contributions to each class of SPT diagrams for Gaussian initial conditions.\label{tab:diagramsGaussian}}
\vspace*{5mm}
\hspace*{30mm}
\begin{tabular}{|r|cccc|}
\hline
  & $B_{1122 \oplus 1113}$  & $B_{xx \otimes yy}$ & $B_{11 \otimes 211}$ & $B_{11 \otimes 11 \otimes 11}$   \\ \hline
 $B_{222}$ & 6  & 0 & 3 & 1  \\
 ${}_{{\rm \MakeUppercase{ \romannumeral 1}}}B_{321}$ & 18  &6 & 18 & 6  \\
 ${}_{{\rm \MakeUppercase{ \romannumeral 2}}}B_{321}$ & 12  &6 & 18 & 6  \\
 $B_{411}$ & 12 & 6 & 15 & 3  \\  
 \hline
\end{tabular}
\end{table}

\begin{table}[t]
\caption{Same as table~\ref{tab:diagramsGaussian}, but for additional contributions arising from non-Gaussian initial conditions.\label{tab:nongauss}}
\vspace*{5mm}
\hspace*{40mm}
\begin{tabular}{|r|ccc|}
\hline
  & $B_{1111 \oplus 1112}$  & $B_{11 \otimes 12}$ & $B_{11 \otimes 111}$    \\ \hline
 ${}_{{\rm \MakeUppercase{ \romannumeral 2}}}B_{112}$ & 3 &0 & 0    \\
 ${}_{{\rm \MakeUppercase{ \romannumeral 1}}}B_{122}$ & 6 &6 & 6   \\
 ${}_{{\rm \MakeUppercase{ \romannumeral 2}}}B_{122}$ & 12 &0 & 3   \\
 ${}_{{\rm \MakeUppercase{ \romannumeral 1}}}B_{113}$ & 6 &0 & 3   \\
 ${}_{{\rm \MakeUppercase{ \romannumeral 2}}}B_{113}$ & 12 &6 & 6   \\
\hline
\end{tabular}
\end{table}

Working with the diagrams in figure~\ref{fig:diagrams}, we find the following  correspondence between the SPT kernels and the LPT correlators.
For Gaussian initial conditions:
\begin{align}
  \tilde{B}_{222}(\fett{k}_1, \fett{k}_2, \fett{k}_3) &=
               -\ii k_{123}C^{(222)}(\fett{k}_1, \fett{k}_2, \fett{k}_3)
               +\frac 1 2 k_{1123}C^{(1122)}(\fett{p}, \fett{k}_1-\fett{p}, \fett{k}_2, \fett{k}_3) \nonumber \\
   &+ \frac 1 2 k_{2213} C^{(1122)}(\fett{p}, \fett{k}_2-\fett{p}, \fett{k}_1, \fett{k}_3) 
               + \frac 1 2 k_{3312}C^{(1122)}(\fett{p}, \fett{k}_3-\fett{p}, \fett{k}_1, \fett{k}_2) \nonumber \\
   &+\ii k_{12}C^{(11)}(\fett{p})\,k_{123}C^{(112)}(\fett{k}_1-\fett{p}, \fett{p}+\fett{k}_2, \fett{k}_3) \nonumber \\
   &
   +\ii k_{13}C^{(11)}(\fett{p})\,k_{123}C^{(121)}(\fett{k}_1-\fett{p}, \fett{k}_2, \fett{p}+\fett{k}_3) \nonumber \\
   &+\ii k_{23}C^{(11)}(\fett{p})\,k_{123}C^{(211)}(\fett{k}_1, \fett{k}_2-\fett{p}, \fett{p}+\fett{k}_3)  \nonumber\\
   &- k_{12}C^{(11)}(\fett{p})\,k_{13}C^{(11)}(\fett{p}-\fett{k}_1)\,k_{23}C^{(11)}(\fett{p}+\fett{k}_2) \,, 
\end{align}   
 \begin{align}  
& {}_{{\rm \MakeUppercase{ \romannumeral 1}}}\tilde{B}_{123}(\fett{k}_1, \fett{k}_2, \fett{k}_3) =
                    -\ii k_{123}\,{ }_{{\rm \MakeUppercase{ \romannumeral 1}}}C^{(123)}(\fett{k}_1, \fett{k}_2, \fett{k}_3) \nonumber \\
  &\hspace{10mm} -\Big[ k_{3312} S_i^{(2)}(\fett{k}_1,\fett{k}_2-\fett{p})\,S_j^{(1)}(-\fett{p})\,S_k^{(1)}(\fett{k}_1)\,S_l^{(2)}(\fett{p},\fett{k}_2-\fett{p})   \Big.
          \nonumber \\
  &\hspace{10mm} +k_{3312} S_i^{(2)}(\fett{k}_1,\fett{p})S_j^{(1)}(\fett{p}-\fett{k}_2)S_k^{(1)}(\fett{k}_1)S_l^{(2)}(\fett{p},\fett{k}_2-\fett{p}) 
   \nonumber \\
  &\hspace{10mm} +k_{3312} S_i^{(1)}(-\fett{p})\,S_j^{(2)}(-\fett{k}_1,\fett{p}-\fett{k}_2)\,S_k^{(1)}(\fett{k}_1)\,S_l^{(2)}(-\fett{p},\fett{p}-\fett{k}_2) 
       \nonumber \\
  &\hspace{10mm} \Big. +k_{3312} S_i^{(1)}(\fett{p}-\fett{k}_2)\,S_j^{(2)}(-\fett{k}_1,-\fett{p})\,S_k^{(1)}(\fett{k}_1)\,S_l^{(2)}(-\fett{p},\fett{p}-\fett{k}_2)  \Big] \nonumber \\
  &\hspace{20mm} \times \pp
      \nonumber \\
  &\hspace{10mm}+\frac 1 2 k_{2213}C^{(1113)}(\fett{p},\fett{k}_2-\fett{p}, \fett{k}_1,\fett{k}_3)+k_{13}C^{(11)}(\fett{k}_1) \,k_{23}C^{(21)}(\fett{k}_2) \nonumber \\
  &\hspace{10mm}+\frac \ii 2 k_{23}C^{(11)}(\fett{p})\, k_{123} C^{(112)}(\fett{k}_1, \fett{k}_2-\fett{p},\fett{p}+\fett{k}_3) \nonumber \\
  &\hspace{10mm}+\frac \ii 2 k_{23}C^{(11)}(\fett{k}_2-\fett{p})\, k_{123} C^{(112)}(\fett{k}_1, \fett{p},-\fett{p}-\fett{k}_1) \nonumber \\
  &\hspace{10mm}+\frac \ii 2 k_{13}C^{(11)}(\fett{k}_1) \left[k_{223}-k_{332} \right] C^{(112)}(\fett{p},\fett{k}_2-\fett{p},-\fett{k}_2) \nonumber \\
  &\hspace{10mm}- \frac 1 2 k_{13}C^{(11)}(\fett{k}_1)\,k_{23}C^{(11)}(\fett{p})\,k_{23}C^{(11)}(\fett{k}_2-\fett{p}) \,, 
\end{align}
\begin{eqnarray}
 {}_{{\rm \MakeUppercase{ \romannumeral 2}}}\tilde{B}_{123}(\fett{k}_1, \fett{k}_2, \fett{k}_3) &=& 
                   -\ii k_{123}\,{ }_{{\rm \MakeUppercase{ \romannumeral 2}}}C^{(312)}(\fett{k}_1, \fett{k}_2, \fett{k}_3) \nonumber \\
  &&-2k_{1123}S_i^{(2)}(-\fett{p},\fett{k}_1)\,S_j^{(1)}(\fett{p})\,S_k^{(1)}(\fett{k}_2)\,S_l^{(2)}(-\fett{k}_1,-\fett{k}_2) \ppp \nonumber \\
  &&-2k_{1123}S_i^{(1)}(\fett{p})\,S_j^{(2)}(\fett{p},-\fett{k}_1)\,S_k^{(1)}(\fett{k}_2)\,S_l^{(2)}(-\fett{k}_1,-\fett{k}_2) \ppp \nonumber  \\
  &&+k_{13}C^{(31)}(\fett{k}_1)\,k_{23}C^{(11)}(\fett{k}_2) +\frac \ii 2 k_{123}C^{(112)}(\fett{k}_1,\fett{k}_2,\fett{k}_3)\,k_{11}C^{(11)}(\fett{p}) \nonumber \\
  &&+k_{23}C^{(11)}(\fett{k}_2)\,k_{113} S_k^{(2)}(-\fett{p},-\fett{k}_1)\,S_l^{(1)}(-\fett{p})\,S_m^{(1)}(-\fett{k}_1)\,P_L(p)\,P_L(k_1) \nonumber \\
  &&+k_{23}C^{(11)}(\fett{k}_2)\,k_{113} S_k^{(1)}(\fett{p})\,S_l^{(2)}(\fett{p},-\fett{k}_1)\,S_m^{(1)}(-\fett{k}_1)\,P_L(p)\,P_L(k_1) \nonumber \\
  &&-\frac 1 2 k_{11}C^{(11)}(\fett{p})\,k_{13}C^{(11)}(\fett{k}_1)\,k_{23}C^{(11)}(\fett{k}_2) \,, 
\end{eqnarray}
\begin{eqnarray}  \label{eq:411}
\tilde{B}_{411}(\fett{k}_1, \fett{k}_2, \fett{k}_3) &=& -\ii k_{123}C^{(114)}(\fett{k}_1,\fett{k}_2,\fett{k}_3)
   +k_{3312}C^{(3111)}(\fett{p},\fett{k}_3-\fett{p},\fett{k}_1,\fett{k}_2) \nonumber \\
  &&-4 k_{3312}S_i^{(2)}(-\fett{k}_1,\fett{p})\,S_j^{(2)}(\fett{k}_2,\fett{p})\,S_k^{(1)}(\fett{k}_1)\,S_l^{(1)}(\fett{k}_2) \,P_L(p)\,P_L(k_1)\,P_L(k_2) \nonumber \\
  &&+k_{13}C^{(11)}(\fett{k}_1)\,k_{23}C^{(13)}(\fett{k}_2)+k_{13}C^{(13)}(\fett{k}_1)\,k_{23}C^{(11)}(\fett{k}_2) \nonumber \\
  &&+\frac \ii 2 k_{123}C^{(112)}(\fett{k}_1,\fett{k}_2,\fett{k}_3)\,k_{33}C^{(11)}(\fett{p}) \nonumber \\
  &&-2k_{13}C^{(11)}(\fett{k}_1)\,k_{332}S_k^{(2)}(\fett{p},-\fett{k}_2)\,S_l^{(1)}(\fett{p})\,S_m^{(1)}(-\fett{k}_2) P_L(p)\,P_L(k_2) \nonumber \\
  &&-2k_{23}C^{(11)}(\fett{k}_2)\,k_{331}S_k^{(2)}(\fett{p},-\fett{k}_1)\,S_l^{(1)}(\fett{p})\,S_m^{(1)}(-\fett{k}_1) P_L(p)\,P_L(k_1) \nonumber \\
  &&-\frac 1 2 k_{33}C^{(11)}(\fett{p})\,k_{13}C^{(11)}(\fett{k}_1)\,k_{23}C^{(11)}(\fett{k}_2)  \,,
\end{eqnarray}
where we have employed the shorthand notation $k_{123} C^{(222)} \equiv k_{1i}k_{2j} k_{3k} C_{ijk}^{(222)}$ and so on,
and omitted writing out the integration over $\fett{p}$.
Note  that these expressions represent only one permutation of the full solution, 
e.g., for the full ${}_{{\rm \MakeUppercase{ \romannumeral 1}}}B_{321}$ one must permute the wavevectors  in
${}_{{\rm \MakeUppercase{ \romannumeral 1}}}\tilde{B}_{123}(\fett{k}_1, \fett{k}_2, \fett{k}_3)$ five times and then sum the six terms together. 
The kernels $\fett{S}^{(a)}$ contain in general both  longitudinal and transverse components.
Contributions to $\tilde{B}$ from transverse components vanish if and only if $\fett{p} \equiv \fett{p}_{12 \cdots a}$ in $\fett{p} \cdot \fett{S}^{(a)}(\fett{p}_1, \ldots, \fett{p}_a)$. As it turns out, the only non-vanishing transverse contribution appears in the second term in equation~(\ref{eq:411}).

For non-Gaussian initial conditions we find:
\be
 {}_{{\rm \MakeUppercase{ \romannumeral 2}}}\tilde{B}_{112}(\fett{k}_1, \fett{k}_2, \fett{k}_3) =
           -\ii k_{123}\,{}_{{\rm \MakeUppercase{ \romannumeral 2}}}C^{(112)}(\fett{k}_1, \fett{k}_2, \fett{k}_3)
           +\frac 1 2 k_{3312}C^{(1111)}(\fett{p}, \fett{k}_3-\fett{p}, \fett{k}_1, \fett{k}_2)   \,,  
\ee
\begin{align}           
 &{}_{{\rm \MakeUppercase{ \romannumeral 1}}}\tilde{B}_{122}(\fett{k}_1, \fett{k}_2, \fett{k}_3) =
           -\ii k_{123}\,{}_{{\rm \MakeUppercase{ \romannumeral 1}}}C^{(122)}(\fett{k}_1, \fett{k}_2, \fett{k}_3) + k_{12}C^{(11)}(\fett{k}_1)\,k_{23}C^{(12)}(\fett{k}_3)  \nonumber \\
  &\hspace{8mm} -k_{3312}S_i^{(1)}(\fett{p})\,S_j^{(1)}(\fett{k}_3-\fett{p})\,S_k^{(1)}(\fett{k}_1)\,P_L(k_1)\,S_l^{(2)}(\fett{k}_1,\fett{k}_3)\,B_0(k_3,p,|\fett{k}_3-\fett{p}|)  \nonumber \\
  & \hspace{8mm}+\frac \ii 2 k_{12}C^{(11)}(\fett{k}_1)\,k_{332}C^{(111)}(\fett{p},\fett{k}_3-\fett{p},-\fett{k}_3)  \,,
 \end{align}
 \begin{align} 
  &{}_{{\rm \MakeUppercase{ \romannumeral 2}}}\tilde{B}_{122}(\fett{k}_1, \fett{k}_2, \fett{k}_3) =
           -\ii k_{123}\,{}_{{\rm \MakeUppercase{ \romannumeral 2}}}C^{(122)}(\fett{k}_1, \fett{k}_2, \fett{k}_3)   \nonumber     \\
   & \hspace{8mm}  - \Big[ 2k_{2213}S_i^{(1)}(\fett{k}_2-\fett{p}) \,P_L(|\fett{k}_2-\fett{p}|)\,S_j^{(1)}(\fett{p})\,S_k^{(1)}(\fett{k}_1)\,S_l^{(2)}(\fett{k}_2-\fett{p},\fett{p}+\fett{k}_1)
               \Big. \nonumber \\ 
   & \hspace{8mm}-2k_{3312}S_i^{(1)}(\fett{p}-\fett{k}_2) \,P_L(|\fett{k}_2-\fett{p}|)\,S_j^{(1)}(-\fett{k}_1-\fett{p})\,S_k^{(1)}(\fett{k}_1)\,S_l^{(2)}(\fett{p}-\fett{k}_2,-\fett{p})
              \nonumber \\ 
   &\hspace{8mm} -\Big. k_{23}C^{(11)}(\fett{k}_2-\fett{p})\,k_{123} S_k^{(1)}(\fett{k}_1)S_l^{(1)}(\fett{p})S_m^{(1)}(-\fett{p}-\fett{k}_1)  \Big] \,B_0(k_1,p,|\fett{k}_1+\fett{p}|)\,, 
  \end{align}
 \begin{align}  
&{}_{{\rm \MakeUppercase{ \romannumeral 1}}}\tilde{B}_{113}(\fett{k}_1, \fett{k}_2, \fett{k}_3) =
           -\ii k_{123}\,{}_{{\rm \MakeUppercase{ \romannumeral 1}}}C^{(311)}(\fett{k}_1, \fett{k}_2, \fett{k}_3) -\frac 1 2 k_{11} C^{(11)}(\fett{p})\,B_0(k_1,k_2,k_3)  \nonumber \\
   & \hspace{15mm}-k_{1123}S_i^{(1)}(\fett{p})\,S_j^{(2)}(\fett{p},-\fett{k}_1)\,S_k^{(1)}(\fett{k}_2)\,S_l^{(1)}(\fett{k}_3)\,P_L(p)\,B_0(k_1,k_2,k_3)  \nonumber \\
   & \hspace{15mm} -k_{1123}S_i^{(2)}(\fett{p},\fett{k}_1)\,S_j^{(1)}(\fett{p})\,S_k^{(1)}(\fett{k}_2)\,S_l^{(1)}(\fett{k}_3)\,P_L(p)\,B_0(k_1,k_2,k_3)  \,, 
\end{align}
\begin{align}
&{}_{{\rm \MakeUppercase{ \romannumeral 2}}}\tilde{B}_{113}(\fett{k}_1, \fett{k}_2, \fett{k}_3) = 
      -\ii k_{123}\,{}_{{\rm \MakeUppercase{ \romannumeral 2}}}C^{(113)}(\fett{k}_1, \fett{k}_2, \fett{k}_3)+ k_{13} C^{(11)}(\fett{k}_1)\,k_{23}C^{(12)}(\fett{k}_2)  \nonumber \\
   & \hspace{15mm}- k_{3312} S_i^{(2)}(\fett{k}_1,\fett{k}_2-\fett{p})
           \,S_j^{(1)}(-\fett{p})\,S_k^{(1)}(\fett{k}_1)\,P_L(k_1)\,S_l^{(1)}(\fett{k}_2)\,B_0 (p,k_2,|\fett{k}_2-\fett{p}|) \nonumber \\
   & \hspace{15mm}-k_{3312}S_i^{(1)}(\fett{p}-\fett{k}_2) \,S_j^{(2)}(\fett{k}_1,\fett{p})\,S_k^{(1)}(\fett{k}_1)\,P_L(k_1)\,S_l^{(1)}(\fett{k}_2) \,B_0 (p,k_2,|\fett{k}_2-\fett{p}|) \nonumber \\
& \hspace{15mm} -\frac \ii 2 k_{13} C^{(11)}(\fett{k}_1)\,k_{332} C^{(111)}(\fett{p},\fett{k}_2-\fett{p},-\fett{k}_2)  \,.
\end{align}
Again, we emphasise that these expressions need to be summed over all their permutations.  In particular, the summation over the three permutations of 
 the term ${}_{{\rm \MakeUppercase{ \romannumeral 2}}}\tilde{B}_{112}(\fett{k}_1, \fett{k}_2, \fett{k}_3)$ is missing in the current SPT literature~\cite{Sefusatti:2010ee,Sefusatti:2009qh}.


\section{Density kernels from LPT\label{appX}}

The symmetrised kernels for equation~(\ref{deltaLag}) are \fleq
\be
  {X}_{1}^{(s)}(\fett{k};\fett{p}_1) =  \fett{k} \cdot \fett{S}^{(1)}(\fett{p}_1)  \,, 
\ee
 \be 
   {X}_{2}^{(s)}(\fett{k};\fett{p}_1,\fett{p}_2)  = \fett{k}\cdot \fett{S}^{(2)}(\fett{p}_1,\fett{p}_2) 
    +\frac 1 2 \fett{k} \cdot \fett{S}^{(1)}(\fett{p}_1) \,\fett{k} \cdot \fett{S}^{(1)}(\fett{p}_2)\,, 
 \ee   
\begin{align}    
  \label{thirdX} {X}_{3}^{(s)}(\fett{k};\fett{p}_1,\fett{p}_2,&\fett{p}_3) = \fett{k} \cdot \fett{S}^{(3)}(\fett{p}_1,\fett{p}_2,\fett{p}_3) 
    + \frac 1 6 \fett{k} \cdot \fett{S}^{(1)}(\fett{p}_1) \,\fett{k} \cdot \fett{S}^{(1)}(\fett{p}_2) \,\fett{k} \cdot \fett{S}^{(1)}(\fett{p}_3) \nonumber \\
    &+  \frac 1 3 \left\{ \fett{k} \cdot \fett{S}^{(1)}(\fett{p}_1)\, \fett{k}\cdot \fett{S}^{(2)}(\fett{p}_2,\fett{p}_3) + \text{two perms.}  \right\} \,, 
\end{align}
\begin{align}    
 \label{fourthX} {X}_4^{(s)}(\fett{k};\fett{p}_1,\fett{p}_2,&\fett{p}_3,\fett{p}_4) =  \fett{k} \cdot \fett{S}^{(4)}(\fett{p}_1,\fett{p}_2,\fett{p}_3,\fett{p}_4) \nonumber \\
          &+\frac{1}{24}   \fett{k} \cdot \fett{S}^{(1)}(\fett{p}_1) \,\fett{k} \cdot \fett{S}^{(1)}(\fett{p}_2) \,\fett{k} \cdot \fett{S}^{(1)}(\fett{p}_3) \,\fett{k} \cdot \fett{S}^{(1)}(\fett{p}_4)   \nonumber \\
 & +\frac 1 3 \Bigg\{\frac 1 2 \Bigg.  \fett{k}\cdot \fett{S}^{(2)}(\fett{p}_1,\fett{p}_2) \,\fett{k}\cdot \fett{S}^{(2)}(\fett{p}_3,\fett{p}_4) +\text{two perms.} \Bigg\}  \nonumber  \\
 &+\frac 1 4 \Bigg\{  \fett{k}\cdot \fett{S}^{(1)}(\fett{p}_1)\,\fett{k} \cdot \fett{S}^{(3)}(\fett{p}_2,\fett{p}_3,\fett{p}_4)  \Bigg. +\text{three perms.} \Bigg\} \nonumber \\
 &+\frac 1 6 \Bigg\{ \frac 1 2 \fett{k}\cdot \fett{S}^{(1)}(\fett{p}_1)\,\fett{k}\cdot 
   \fett{S}^{(1)}(\fett{p}_2)\, \fett{k}\cdot \fett{S}^{(2)}(\fett{p}_3,\fett{p}_4) +\text{five perms.}  \Bigg\}  \,. 
\end{align}
At this point it is important to ask \emph{when}  purely transverse Lagrangian modes affect the $n$th order density contrast. The first term
on the RHS of equation~(\ref{thirdX}) is $\fett{k} \cdot \fett{S}^{(3)}(\fett{p}_1,\fett{p}_2,\fett{p}_3)$, 
where the $\fett{S}^{(3)}$ kernels contains a transverse term $\propto \fett{\omega}_{3c}^{(s)}(\fett{p}_1,\fett{p}_2,\fett{p}_3)$ (see equation~(\ref{ls3rd})). 
At third order, the Dirac delta in equation~(\ref{deltaLag}) fixes $\fett{k} = \fett{p}_{123}$, thus leading to $\fett{p}_{123} \cdot \fett{\omega}_{3c}^{(s)}(\fett{p}_1,\fett{p}_2,\fett{p}_3) = 0$ because of the transverseness condition~(\ref{transverseCondition}), i.e.,
the third order transverse kernel $\fett{\omega}_{3c}^{(s)}$ does not contribute to  $\tilde\delta^{(3)}$. However, when a similar $\fett{k} \cdot \fett{S}^{(3)}(\fett{p}_1,\fett{p}_2,\fett{p}_3)$
term appears in the $\tilde\delta^{(4)}$ expression~(\ref{fourthX}), the Dirac delta now implies $\fett{k}=\fett{p}_{1234}$, so that  $\fett{p}_{1234} \cdot \fett{\omega}_{3c}^{(s)}(\fett{p}_1,\fett{p}_2,\fett{p}_3) =\fett{p}_{4} \cdot \fett{\omega}_{3c}^{(s)}(\fett{p}_1,\fett{p}_2,\fett{p}_3)$, which is generally non-vanishing.  Hence, the transverse term $\fett{\omega}_{3c}^{(s)}$ does contribute to  $\tilde\delta^{(4)}$, and we can conclude  by induction that all $n$th order transverse kernels must contribute to 
 $\tilde\delta^{(N>n)}$, but not $\tilde\delta^{(N\leq n)}$.


\end{document}